\documentclass{article}

\PassOptionsToPackage{numbers}{natbib}
\usepackage[preprint]{neurips_2024}

\usepackage[utf8]{inputenc} % allow utf-8 input
\usepackage[T1]{fontenc}    % use 8-bit T1 fonts
\usepackage{hyperref}       % hyperlinks
\usepackage{url}            % simple URL typesetting
\usepackage{booktabs}       % professional-quality tables
\usepackage{amsfonts}       % blackboard math symbols
\usepackage{nicefrac}       % compact symbols for 1/2, etc.
\usepackage{microtype}      % microtypography
\usepackage{xcolor}         % colors

\usepackage[flushleft]{threeparttable}
\usepackage{multicol,multirow}
\usepackage{amsmath}
\usepackage{amssymb}
\usepackage{graphicx}
\usepackage{array}
\usepackage[numbers]{natbib}
\newcolumntype{M}[1]{>{\centering\arraybackslash}m{#1}}
\usepackage{lipsum}
\title{DEX-TTS: Diffusion-based EXpressive Text-to-Speech with Style Modeling on Time Variability}

\author{%
  Hyun Joon Park, Jin Sob Kim, Wooseok Shin, Sung Won Han\thanks{Corresponding author.}  \\
  Korea University\\
  \texttt{\{winddori2002, jinsob, wsshin95, swhan\}@korea.ac.kr} \\
}

\begin{document}

\maketitle

\begin{abstract}
Expressive Text-to-Speech (TTS) using reference speech has been studied extensively to synthesize natural speech, but there are limitations to obtaining well-represented styles and improving model generalization ability. In this study, we present Diffusion-based EXpressive TTS (DEX-TTS), an acoustic model designed for reference-based speech synthesis with enhanced style representations. Based on a general diffusion TTS framework, DEX-TTS includes encoders and adapters to handle styles extracted from reference speech. Key innovations contain the differentiation of styles into time-invariant and time-variant categories for effective style extraction, as well as the design of encoders and adapters with high generalization ability. In addition, we introduce overlapping patchify and convolution-frequency patch embedding strategies to improve DiT-based diffusion networks for TTS. DEX-TTS yields outstanding performance in terms of objective and subjective evaluation in English multi-speaker and emotional multi-speaker datasets, without relying on pre-training strategies. Lastly, the comparison results for the general TTS on a single-speaker dataset verify the effectiveness of our enhanced diffusion backbone. Demos are available here.\footnote{Audio samples are available at \url{https://dextts.github.io/demo.github.io/}}
\end{abstract}

\section{Introduction}
\label{sec:intro}

Text-to-Speech (TTS) \cite{oord2016wavenet, wang2017tacotron, shen2018natural, ren2019fastspeech} is the task of synthesizing natural speech from a given text, which is applied to various applications such as voice assistant services. To generate diverse and high-fidelity speech, researchers have studied Transformer \cite{li2019neural}-, GAN \cite{yamamoto2020parallel, yang2021ganspeech}-, and normalizing flow \cite{miao2020flow, kim2020glow}-based TTS as deep generative models. Recently, with the success of diffusion models in various generative tasks \cite{ho2020denoising, ho2022imagen, schneider2023archisound}, researchers have shifted their focus to diffusion-based TTS \cite{kong2020diffwave, jeong2021diff, popov2021grad, liu2022diffgan} and proved the outstanding performance of diffusion models in TTS as well.

Despite the improvement of the above general TTS studies, synthesizing human-like speech remains challenging due to the lack of expressiveness of synthesized speech such as the limited styles of reading, prosody, and emotion \cite{tan2021survey}. Since expressiveness can be reflected during synthesizing acoustic features such as mel-spectrograms, acoustic models have been investigated for expressive TTS. Although some studies \cite{lee2017emotional, kim2021expressive, li2021controllable} generate expressive speech using emotion labels or style tags, the necessity of label information constrains the applicability of the methods.

Considering the previous limitation, researchers have adopted reference-based TTS, which can operate without explicit labels \cite{wang2018style, valle2020mellotron, lee2021styler, min2021meta, casanova2022yourtts, li2022styletts, huang2022generspeech}, for expressive TTS. These methods extract styles (e.g., emotion, timbre, and prosody) from reference speech and reflect these styles in the speech. For real-world applications, the reference-based TTS is designed to enable handling unseen reference speech, like reference speech from unseen speakers during training.

As aforementioned, expressive TTS utilizing a reference involves two steps, extracting the reference information (extractor) and incorporating the information into the synthesis process (adapter). For outstanding expressive TTS, the extractor and adapter should be designed based on the following two aspects: a well-represented style and generalization. That is, expressive TTS should extract rich styles from references and incorporate these styles into the synthesis process. Furthermore, it should have a high generalization ability to operate even in zero-shot scenarios. However, previous studies lacked considerations for network design from the above perspectives, resulting in lower performance in zero-shot or insufficient style reflection. It can be exacerbated when expressive speech is used as a reference because expressive speech contains diverse style information. It suggests the necessity of the network design under the above perspectives for expressive TTS. Some studies \cite{casanova2022yourtts, li2022styletts, huang2022generspeech} attempted to address this problem through pre-training stages or networks. However, problems such as complicated pipelines, additional label requirements, and dependencies on other models remain.

Another focus of this study is designing a strong TTS backbone, which is a component of expressive TTS, to obtain superior expressive TTS. We investigate diffusion-based TTS since it can synthesize high-quality speech through iterative denoising processes. Furthermore, we expect that style information can be effectively reflected by iteratively incorporating style information during the denoising process. A few studies \cite{jeong2021diff, popov2021grad, ye2023comospeech} on diffusion-based TTS have improved TTS performance by adapting diffusion formulations to suit TTS. However, the network of these studies was confined to simple U-Net, leading to limited latent representations. Although U-DiT-TTS \cite{jing2023u} used DiT \cite{peebles2023scalable} for the diffusion network, DiT is yet to be effectively leveraged for TTS.

To address the discussion, we propose a novel acoustic model, Diffusion-based EXpressive TTS (DEX-TTS). Based on a general diffusion TTS, DEX-TTS contains encoders and adapters to handle the styles of reference speech. First, we adopt overlapping patchify and convolution-frequency patch embedding to enhance the DiT-based diffusion TTS backbone. Furthermore, we separate styles into time-invariant and time-variant styles to extract diverse styles even from expressive reference speech. We design each time-invariant and time-variant encoder which utilizes multi-level feature maps and vector quantization, making well-refined styles. Lastly, we propose time-invariant and time-variant adapters that incorporate each extracted style into the speech synthesis process. For effective style reflection and high generalization capability, each method is based on Adaptive Instance Normalization (AdaIN) \cite{huang2017arbitrary} and cross-attention \cite{vaswani2017attention} methods. To effectively leverage the iterative denoising process of diffusion TTS, we design adapters that adaptively reflect styles over time. Through the proposed methods, we can synthesize high-quality and reference-style speech. 

We conduct experiments on multi-speaker and emotional multi-speaker datasets to verify the proposed methods. The results reveal that, including zero-shot scenarios, DEX-TTS achieves more outstanding performance than the previous expressive TTS methods in terms of speech quality and similarity. Unlike some existing methods that rely on pre-training strategies, DEX-TTS achieves superior performance as an independent model. Furthermore, to investigate the effect of our strategies to improve the diffusion TTS backbone, we conduct experiments on general TTS using our diffusion backbone. The results on the single-speaker dataset demonstrate the superior performance of the proposed method compared with previous diffusion TTS methods.

\section{Related Works}
\label{sec:rel}

\subsection{Diffusion-based Text-to-Speech}

As diffusion models in image synthesis have proven their outstanding performance \cite{ho2020denoising, song2020score, karras2022elucidating}, researchers have studied diffusion-based TTS. Diff-TTS \cite{jeong2021diff}, Grad-TTS \cite{popov2021grad}, and CoMoSpeech \cite{ye2023comospeech} properly utilized diffusion methods for TTS. Although they effectively applied diffusion formulation for TTS, diffusion networks in previous studies were limited to U-Net architectures. It led to limited latent representations, indicating the necessity of improvement in the diffusion network design. U-DIT-TTS \cite{jing2023u} improved the design of diffusion networks in TTS by adopting DiT \cite{peebles2023scalable} blocks while retaining some aspects of U-Net down- and up-sampling. DiT can extract detailed representations using attention operations between small patches. However, U-DiT-TTS applied large patch strategies and sinusoidal position embedding to synthesize speech of variable time lengths, preventing it from fully leveraging the advantages of DiT. In our work, we adopt an overlapping patchify and convolution-frequency patch embedding to exploit the advantage of DiT structure fully and to design an improved diffusion-based TTS model.

\subsection{Expressive Text-to-Speech}

Reference-based expressive TTS has attracted considerable interest due to the limitations of previous studies that require additional label information \cite{lee2017emotional, kim2021expressive, li2021controllable}. To condition reference information, some studies \cite{wang2018style, valle2020mellotron, casanova2022yourtts} used summation or concatenation, but these methods exhibited limited performance in zero-shot. On the other hand, MetaStyleSpeech \cite{min2021meta} and StyleTTS \cite{li2022styletts} utilized adaptive normalization as a style conditioning method for robust performance in zero-shot. However, in these methods, pooling was applied to reference representations to obtain only a single-style vector, which did not effectively extract diverse styles from references. Although GenerSpeech \cite{huang2022generspeech} proposed a multi-level style adapter to obtain diverse styles, their conditioning method during the synthesis process was confined to summation or concatenation. Previous studies lacked in designing methods to effectively process styles and improve generalization ability. Furthermore, previous studies \cite{casanova2022yourtts, li2022styletts, huang2022generspeech} have limitations requiring pre-training strategies for feature extraction. In our work, we introduce a novel standalone diffusion-based TTS that handles well-represented styles with dedicated extractors and adapters and exhibits strong generalization performance.

\section{DEX-TTS}

\subsection{Preliminaries}
\label{subsec:Pre}

\paragraph{Diffusion Formulation} 
Before introducing our methods, we review the diffusion used in the study. The diffusion model consists of two processes: the diffusion process, which adds Gaussian noise to the original data, and the reverse process, which removes Gaussian noise to generate samples. Diffusion process yields noisy data $\{x\}_{t=0}^{T}$, where $p_{0}(x)$ is a data distribution $p_{data}(x)$ and $p_{T}(x)$ is indistinguishable from pure Gaussian noise. Song et al. \cite{song2020score} addressed the diffusion process as a stochastic process over time $t$ and defined diffusion process with Stochastic Differential Equations (SDE) as $dx = f(x,t)dt + g(t)dw$ where $w$ is Brownian motion, and $f(\cdot,t)$ and $g(\cdot)$ are drift and diffusion coefficients. Song et al. \cite{song2020score} also presented that a probability flow Ordinary Differential Equations (ODE) corresponds to the deterministic process of SDE, and it is defined as below:
%###########################
\begin{equation}\label{eq:SDE}
  \begin{gathered}
    dx = [f(x,t)-\frac{1}{2}g(t)^{2}\triangledown_{x}logp_{t}(x)]dt
  \end{gathered}
\end{equation}
%###########################
The deterministic process is determined from the SDE when the score predicted by the score function $\triangledown_{x}logp_{t}(x)$ is known. For the reverse process, a numerical ODE solver such as Euler can be used. 

EDM \cite{karras2022elucidating} defines the score function as $\triangledown_{x}logp_{t}(x)=(D_{\theta}(x,t)-x_{t})/\sigma^{2}_{t}$ given $\sigma^{2}_{t}$ is $\int g(t)^{2}dt$, and $D_{\theta}$ is a denoiser network trained by denoising error $||D_{\theta}(x_{t},t)-x||_{2}^{2}$. To train a denoiser while avoiding gradient variation, EDM introduces pre-conditioning and $t$-dependent skip connection which are also investigated in CoMoSpeech \cite{ye2023comospeech} where the schedule $\sigma_{t}$ is $t$.
%###########################
\begin{equation}\label{eq:EDM}
  \begin{gathered}
    D_{\theta}(x_{t},t)=c_{skip}(t)x_{t} + c_{out}(t)F_{\theta}(c_{in}(t)x, c_{noise}(t))
  \end{gathered}
\end{equation}
%##########################
$F_{\theta}$ is the network before conditioning. We follow Equation \ref{eq:EDM} with the parameter settings in \cite{karras2022elucidating} to build diffusion. In practice, we can forward text and style representations into $D_{\theta}$ and $F_{\theta}$ to condition the denoising process, as described in \cite{ye2023comospeech}.

\begin{figure*}[t]
  \centering
  \includegraphics[scale=0.59]{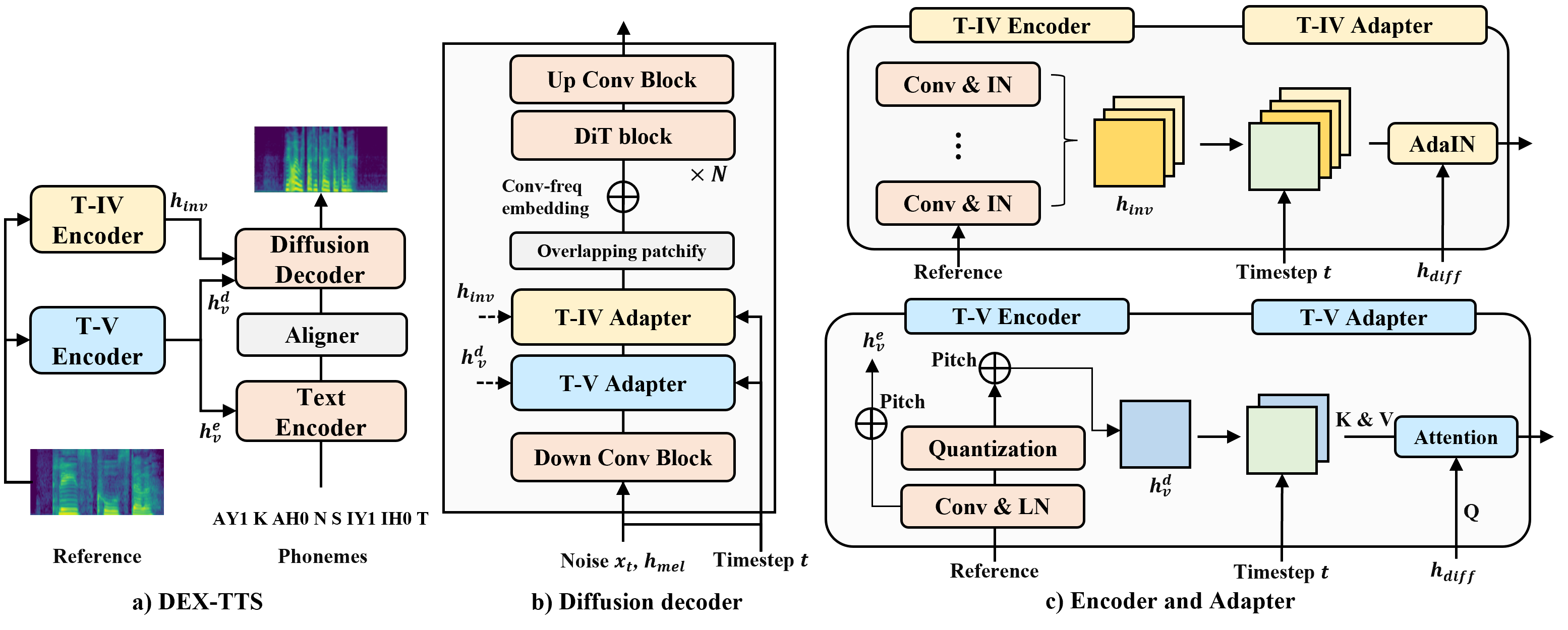} 
  \caption{Architecture of DEX-TTS, diffusion decoder, and style encoders and adapters.} 
  \label{fig:dextts}
\end{figure*}

\subsection{Overall Architecture} 
Figure \ref{fig:dextts} depicts the architecture of DEX-TTS. In our TTS system, a vocoder is applied to the synthesized mel-spectrogram to convert it into a signal. DEX-TTS contains encoders and adapters to extract and incorporate style information based on a basic TTS architecture. To effectively extract styles from the reference speech, we define style information as two features, namely time-invariant (T-IV) and time-variant (T-V) styles. T-IV styles contain global information that rarely varies within speech, whereas T-V styles contain information that varies within speech, such as intonation. Based on this approach, the T-IV encoder passes extracted representation $h_{inv}$ to the diffusion decoder, and the T-IV adapter reflects it regardless of the time axis. On the other hand, to preserve the temporal information of the representation $h^{d}_{v}$ from the T-V encoder, the T-V adapter in the diffusion decoder reflects the representation by considering the time axis. In addition, the T-V encoder forwards the representation $h^{e}_{v}$ to the text encoder since the text representation varies over time.

\paragraph{Text Encoder} 
Given the input phonemes, the text encoder extracts the text representation $h_{text}$. The text encoder consists of 8 layers, each composed of Transformer encoder structure \cite{vaswani2017attention} that includes Multi-Head Self-Attention (MHSA) and Feed-Forward Network (FFN). To enhance the encoder, we incorporate relative position embedding, RoPE \cite{su2024roformer}, into the attention mechanism and apply the swish gate used in RetNet \cite{sun2023retentive} after the attention operation. Since text varies over time, $h^{e}_{v}$, defined as T-V styles, can be effectively utilized to condition styles for $h_{text}$. We utilize Adaptive Layer Normalization (AdaLN) \cite{chen2021adaspeech} after each MHSA and FFN to inject $h^{e}_{v}$, extracted by the T-V encoder given a reference input, into $h_{text}$. See Section \ref{sec:appendix-text encoder} for more details.

\paragraph{Aligner} 
We adopt a convolution-based Duration Predictor (DP) \cite{kim2020glow} in which $h_{text}$ extracted by the text encoder is used. DP predicts the duration $\hat{d}$ which maps $h_{text}$ to frames of the mel-spectrogram for the initial mel-spectrogram representation $h_{mel}$ used as the condition input in the diffusion decoder. Aligner is trained by the Monotonic Alignment Search (MAS) algorithm.

\paragraph{Diffusion Decoder} 
Given time $t$ and corresponding noise $x_{t}$ generated by the diffusion process, the diffusion decoder synthesizes a denoised mel-spectrogram $\hat{x}$. Here, the initial mel-spectrogram representation $h_{mel}$ and styles $h_{inv, v}$ are utilized as conditioning information. For diffusion representation $h_{diff}$, we concatenate $x_{t}$, $h_{mel}$, and $t$, and pass it to the diffusion decoder, where $t$ is projected by sinusoidal encoding and linear layers.

The diffusion decoder comprises convolution blocks, adapters, and DiT blocks \cite{peebles2023scalable}. To leverage powerful denoising in the latent space, we utilize up and down convolution blocks, as in \cite{popov2021grad}, to decrease and increase the resolution of $h_{diff}$. In the bottleneck, each adapter incorporates $h_{inv, v}$ into $h_{diff}$ to reflect each style information. Furthermore, we forward $t$ as an additional condition into adapters to effectively reflect styles during the iterative denoising process.

After adapters, we utilize DiT blocks to enhance latent representations. To effectively exploit DiT blocks, we introduce overlapping patchify and convolution-frequency (conv-freq) patch embedding. Unlike previous methods, we allow overlapping between patches, mitigating boundary artifacts between patches and enabling natural speech synthesis. For patchify, a convolution layer with a kernel size of $2 \times P - 1$ and stride of $P$ is used given patch size $P$. Let $h \in \mathbb{R}^{C \times F \times T}$ be the representations after adapters, where $C$, $F$, and $T$ are the hidden, frequency, and time sizes respectively. Given $F_{2}=F/P$ and $T_{2}=T/P$, the convolution layer patchifies $h$ into $h_{p} \in \mathbb{R}^{C \times F_{2} \times T_{2}}$.

Before converting the spatial dimensions of $h_{p}$ into a sequence for DiT inputs, embeddings are added to each frequency and time dimension. Since the speech length is variable, patch embedding should be able to handle unseen lengths during training. For the time axis, we apply a convolution layer to $h_{p}$ and take a time-wise average to obtain the relative positional embeddings  $PE_{T} \in \mathbb{R}^{C \times 1 \times T_{2}}$. On the other hand, we use fixed-size learnable parameters as frequency embedding $PE_{F} \in \mathbb{R}^{C \times F_{2} \times 1}$ since the frequency size is not variable in speech synthesis. $PE_{T}$ and $PE_{F}$ are added to $h_{p}$, and the spatial dimension is converted into a sequence to be used as input for DiT. By the above embedding approach, we can get robust embedding for variable lengths compared to conventional embeddings obtained by fixed-size parameters or sinusoidal encoding. After the DiT block, the up-convolution block predicts a denoised mel-spectrogram from latent features.

\subsection{Time-Invariant Style Modeling} 

We model T-IV and T-V styles to extract well-represented styles from the reference speech. For T-IV styles, we design the encoder and adapter to process global information within the speech.

\paragraph{T-IV Encoder} 
As depicted in Figure \ref{fig:dextts}.c, the T-IV encoder consists of a few residual convolution blocks to extract the representation from the reference speech. To maintain individual characteristics regardless of the temporal information within a batch, we utilize Instance Normalization (IN) after each block. Inspired by \cite{park2023triaan}, we use multi-level feature maps as T-IV styles $ h_{inv} \in \mathbb{R}^{L \times C \times T}$, where $L$ is the number of layers in the T-IV encoder. Since $h_{inv}$ comprises all stacked feature maps from the convolution block, it contains T-IV information across the convolution blocks.

\paragraph{T-IV adapter} 
For T-IV Adatpor, we apply AdaIN, which can reflect styles regardless of temporal information, to inject $h_{inv}$ into $h_{diff}$ during the denoising process in the diffusion decoder. Given mean $\mu$ and standard deviation $\sigma$ as bias and scale for AdaIN, the process is defined as follows:
%###########################
\begin{equation}\label{eq:AdaIN}
  \begin{gathered}
    AdaIN(h_{diff},\mu, \sigma) = IN(h_{diff}) \times \sigma + \mu
  \end{gathered}
\end{equation}
%##########################
where $\mu$ and $\sigma$ are obtained using $h_{inv}$. Since $h_{inv}$ contains the feature maps of all layers, we compute the channel-wise mean and standard deviation for each layer and utilize attention pooling (AP) \cite{park2023triaan} to obtain representative statistics ($\mu$ and $\sigma$) for AdaIN. Furthermore, we include time $t$ in the pooling process to ensure adaptive operation at each time step during the denoising process. The process for extracting $\mu$ and $\sigma$ is represented as follows:
%###########################
\begin{equation}\label{eq:attention pooling}
  \begin{gathered}
    AP(x)=sum(softmax(xW_{ap}) \times x) \\
    \tilde{\mu}=[t; avg(h^{1}_{inv}); ..., ;avg(h^{L}_{inv})], \ \mu = AP(\tilde{\mu}) \\
    \tilde{\sigma}=[t; std(h^{1}_{inv}); ..., ;std(h^{L}_{inv})], \ \sigma = AP(\tilde{\sigma}) \\
  \end{gathered}
\end{equation}
%##########################
where $W_{ap}$ is the linear weight for AP. Through AP, we can extract common features across multi-level feature maps to utilize as a general T-IV style. Furthermore, time conditioning enables adaptive style incorporation at each timestep.

\subsection{Time-Variant Style Modeling}

We define T-V styles as features that emerge with temporal variation within speech. Based on this, we design the encoder and adapter to preserve or reflect the temporal information of the reference.

\paragraph{T-V Encoder} 
Similar to the T-IV encoder, the T-V encoder contains a few residual convolution blocks, but we employ Layer Normalization (LN) instead of IN to preserve temporal relationships in each instance. The T-V encoder extracts two styles $h^{e,d}_{v}$ for each text encoder and diffusion decoder. We obtain $h^{e}_{v}$ by applying convolution blocks to the reference and adding the pitch information $h_{f0}$. We use $h_{f0}$ to reflect changes in speech over time, and it is extracted by applying GRU layers to the log fundamental frequency of the reference speech. After channel-wise pooling on $h^{e}_{v}$ to obtain the overall T-V style information, $h^{e}_{v}$ is forwarded to the text encoder.

For $h^{d}_{v}$, we additionally apply Vector Quantization (VQ) to the output of convolution blocks. For VQ, we utilize a latent discrete codebook $e \in \mathbb{R}^{K \times D}$, where $K$ is the codebook size and $D$ is the dimension size. The VQ layer maps the outputs to a discrete space based on the distance to the codebook. This process removes noise from a continuous space, obtaining well-refined style information that can be used as generalized features. After adding the pitch information, $h^{d}_{v}$ is passed to the decoder without pooling to preserve temporal information.

\paragraph{AdaLN} 
In the text encoder, we apply AdaLN to reflect the overall style $h^{e}_{v}$ while preserving the temporal aspects of the text representation. AdaLN with $h_{text}$ in the encoder is defined as follows:
%###########################
\begin{equation}\label{eq:Adaln}
  \begin{gathered}
    AdaLN(h_{text},h^{e}_{v})=g(h^{e}_{v})\times LN(h_{text}) + b(h^{e}_{v})
  \end{gathered}
\end{equation}
%##########################
where $g(\cdot)$ and $b(\cdot)$ are linear layers for scaling and bias, respectively.

\paragraph{T-V adapter} 
To reflect T-V styles $h^{d}_{v}$ to $h_{diff}$ while preserving temporal information, we design the T-V adapter with cross-attention. We use $h_{diff}$ as the query and $h^{d}_{v}$ as the key and values for cross-attention (CA), and it is defined as below.
%###########################
\begin{equation}\label{eq:t-v adapter}
  \begin{gathered}    
    Q=IN(h_{diff})W_{q},\ \ K=h^{d}_{v}W_{k}, \ \ V=h^{d}_{v}W_{v} \\ 
    CA(Q,K,V)=softmax(QK^{\top})V
  \end{gathered}
\end{equation}
%##########################
where $W_{q,k,v}$ denotes the linear weight. As presented in Equation \ref{eq:t-v adapter}, IN is applied to $h_{diff}$ for the query. It maintains instance-level features for computing attention scores and enables reflecting suitable T-V style for each instance.

\subsection{Loss Function} 
To train DEX-TTS, we follow the loss formulation of previous diffusion-based TTS studies \cite{popov2021grad, ye2023comospeech}, in which duration loss $\mathcal{L}_{dur}$, prior loss $\mathcal{L}_{prior}$, and diffusion loss $\mathcal{L}_{diff}$ are used. $\mathcal{L}_{dur}$ is utilized to train DP that predicts the duration mapping $h_{text}$ to mel frames, and it is defined as $||log(d)-log(\hat{d})||_{2}^{2}$. The duration label $d$ is obtained by MAS algorithm. Given $x$ is the ground-truth mel-spectrogram, $\mathcal{L}_{prior}$ calculates the loss between the initial mel-spectrogram $h_{mel}$ from the aligner and $x$ for stable learning, defined as $||h_{mel}-x||_{2}^{2}$. To train the diffusion decoder (Denoiser $D_{\theta}$), we use a denoising error for each timestep $t$, defined as follows:
%###########################
\begin{equation}\label{eq:Diff loss}
  \begin{gathered}
    \mathcal{L}_{diff}=\lambda(t)||D_{\theta}(x_{t},t, h_{\{mel, inv, v\}})-x||_{2}^{2}
  \end{gathered}
\end{equation}
%##########################
$\lambda(t)$ is the weight for noise levels determined by $t$ used in \cite{karras2022elucidating}. For VQ loss, we adopt a commitment loss \cite{van2017neural} $\mathcal{L}_{vq}$ defined as $||h-sg(e)||_{2}^{2}$, where $h$ is the representation before quantization and $sg$ is the stop-gradient operation. We get the total loss $\mathcal{L}$ by the summation of $\mathcal{L}_{dur}$, $\mathcal{L}_{prior}$, $\mathcal{L}_{diff}$, and $\mathcal{L}_{vq}$.

\section{Experiments}

\subsection{Experiment Setup}
\label{sec:experiment setup}

\paragraph{Dataset} 
To evaluate the proposed method, we use the VCTK dataset \cite{vctk2019}, an English multi-speaker dataset, consisting of approximately 400 utterances per 109 speakers. We split the dataset into about 70\%, 15\%, and 15\% for the train, validation, and test sets, respectively, based on speakers to consider both the seen and unseen (zero-shot) scenarios. For the zero-shot scenario, 10 unseen speakers are used. In addition, we conduct experiments on the Emotional Speech Dataset (ESD) \cite{zhou2021seen} to verify whether the models can reflect styles using expressive reference speech. The ESD contains 10 English and 10 Chinese speakers with 400 sentences per speaker for five emotions (happy, sad, neutral, surprise, and angry). We only use English speakers and keep the same split ratio as that in the VCTK dataset. Two unseen speakers are used for the zero-shot scenario. Considering real-world applications, we design both parallel and non-parallel test scenarios based on whether the input text is the same as the text of the reference speech. For the experimental results, we record the average performances of the parallel and non-parallel scenarios. Finally, all datasets are resampled to 22 kHz.

\paragraph{Baselines} 
For comparison, we set the following systems as baselines: \textbf{1) Ref}, the reference audio. \textbf{2) MetaStyleSpeech} \cite{min2021meta}, multi-speaker adaptive TTS  with meta-learning. \textbf{3) YourTTS} \cite{casanova2022yourtts}, VITS-based zero-shot multi-speaker TTS with the pre-trained speaker encoder. \textbf{4) GenerSpeech} \cite{huang2022generspeech}, style transfer method for out-of-domain TTS. \textbf{5) StyleTTS} \cite{li2022styletts}, style-based TTS with transferable aligner and AdaIN. Except for YourTTS (end-to-end TTS), the generated mel-spectrograms are transformed into waveforms by the pre-trained  HiFi-GAN \cite{kong2020hifi}. We record the performance of the baselines after training with their codes.

\paragraph{Implementation Details} 
For training, we take 1000 and 1500 epochs for the VCTK and ESD datasets, respectively. An Adam optimizer with a learning rate of $10^{-4}$ and batch size of 32 are used. Regarding the model hyperparameters of the diffusion decoder, we take a patch size $P$ of 2, number of DiT blocks $N$ of 4, and hidden size $C$ of 64. The T-IV and T-V encoders use the 6 layers $L$, and their dimension sizes are matched with the diffusion decoder. We set a codebook size $K$ of 512 and dimension size $D$ of 192 for the VQ layer in the T-V encoder. We extract mel-spectrograms with 80 mel bins based on the FFT size of 1024, hop size of 256, and window size of 1024, which is compatible with the HiFi-GAN vocoder used in our TTS system. For the diffusion denoising steps, we use 50 Number of Function Evaluations (NFE) with the Euler solver (See Section \ref{sec:nfe and rtf} to find results depending on NFE). All experiments are conducted on a single NVIDIA 3090 GPU. Codes are available here.\footnote{Codes are available at \url{https://github.com/winddori2002/DEX-TTS/}}

%%%%%%%%%%%%%%%%%%%%%%%%%%%%%%%
\begin{table}
\caption{Comparison results for expressive TTS on the VCTK dataset.}\label{tab:exp-vctk}
\centering
\resizebox{1\columnwidth}{!}{%
\begin{tabular}{p{3.5cm}M{1.2cm}M{1.2cm}M{1.2cm}M{1.2cm}cM{1.2cm}M{1.2cm}M{1.2cm}M{1.2cm}}
% \begin{tabular}{p{2.5cm}ccccccccccc}
\toprule
\multirow{2.5}{*}{Model} & \multicolumn{4}{c}{Seen scenarios} & & \multicolumn{4}{c}{Unseen (zero-shot) scenarios} \\ 
\cmidrule{2-5} \cmidrule{7-10} & WER & COS & MOS-N & MOS-S & & WER & COS & MOS-N & MOS-S \\
\midrule

Ref                                      & 6.23  & -      & 3.97 & -    & & 6.23  & -      & 3.97 & -  \\
MetaStyleSpeech \cite{min2021meta}       & 16.58 & 78.10  & 3.43 & 3.63 & & 16.50 & 73.53  & 3.38 & 3.30  \\
YourTTS \cite{casanova2022yourtts}       & 21.27 & 78.78  & 3.20 & 3.11 & & 18.34 & 75.00  & 3.33 & 3.08  \\
GenerSpeech \cite{huang2022generspeech}  & 13.87 & 77.46  & 3.40 & 3.25 & & 11.37 & 73.23  & 3.46 & 3.06  \\
StyleTTS \cite{li2022styletts}           & \textbf{7.72}  & 82.93  & 3.57 & 3.70 & & 6.58  & 77.90  & 3.53 & 3.65  \\

\midrule
DEX-TTS (ours)                           & 7.85  & \textbf{85.31} & \textbf{3.75} & \textbf{3.88} & & \textbf{5.84}  & \textbf{80.45} & \textbf{3.76} & \textbf{3.81} \\
\bottomrule
\end{tabular}}
\end{table}
% %%%%%%%%%%%%%%%%%%%%%%%%%%%%%%%%

%%%%%%%%%%%%%%%%%%%%%%%%%%%%%%%
\begin{table}
\caption{Comparison results for expressive TTS on the ESD dataset.}\label{tab:exp-esd}
\centering
\resizebox{1\columnwidth}{!}{%
\begin{tabular}{p{3.5cm}M{1.2cm}M{1.2cm}M{1.2cm}M{1.2cm}cM{1.2cm}M{1.2cm}M{1.2cm}M{1.2cm}}
% \begin{tabular}{p{2.5cm}ccccccccccc}
\toprule
\multirow{2.5}{*}{Model} & \multicolumn{4}{c}{Seen scenarios} & & \multicolumn{4}{c}{Unseen (zero-shot) scenarios} \\ 
\cmidrule{2-5} \cmidrule{7-10} & WER  & COS & MOS-N & MOS-S & & WER & COS & MOS-N & MOS-S \\
\midrule

Ref                                      & 7.12  & -      & 3.90 & -    & & 7.12  & -      & 3.90 & -  \\
MetaStyleSpeech \cite{min2021meta}       & 24.56 & 79.41  & 3.09 & 3.53 & & 25.84 & 73.01  & 3.19 & 3.34  \\
YourTTS \cite{casanova2022yourtts}       & 16.57 & 77.61  & 3.33 & 3.40 & & 16.35 & 69.38  & 3.28 & 2.96  \\
GenerSpeech \cite{huang2022generspeech}  & 12.75 & 75.09  & 3.23 & 3.28 & & 11.78 & 70.54  & 3.06 & 2.78  \\
StyleTTS \cite{li2022styletts}           & 12.59 & 79.65  & 3.41 & 3.50 & & 12.11 & 72.27  & 3.23 & 3.06  \\
\midrule
DEX-TTS (ours)                           & \textbf{8.34}  & \textbf{82.71} & \textbf{3.73} & \textbf{3.84} & & \textbf{8.35} & \textbf{75.58} & \textbf{3.57} & \textbf{3.52} \\
\bottomrule
\end{tabular}}
\end{table}
% %%%%%%%%%%%%%%%%%%%%%%%%%%%%%%%%

\paragraph{Evaluation Metrics} 
We consider objective and subjective evaluation metrics. As objective metrics, we utilize Word Error Rate (WER \%) and Cosine Similarity (COS). WER represents how accurately the model synthesizes the given text, and it is calculated as the error between the predicted text, obtained by applying the pre-trained Wav2Vec 2.0 \cite{baevski2020wav2vec} to the synthesized speech, and the given text. On the other hand, COS indicates the similarity in the feature space between the synthesized and reference speech, and it is calculated using a pre-trained speaker verification model.\footnote{https://github.com/resemble-ai/Resemblyzer} For convenience, we show COS multiplied by 100 in the experimental results. For the subjective metrics, we adopt Mean Opinion Score for naturalness and similarity (MOS-N and S). We use Amazon Mechanical Turk (AMT) and ask participants to score on a scale from 1 to 5. They assess the synthesized speech for its naturalness by listening to it, or they compare the synthesized speech with reference speech to evaluate similarity. For every MOS evaluation, we randomly select 30 utterances for each model and guarantee at least 27 participants.

\subsection{Experimental Results}

We conduct experiments including seen and unseen scenarios on multi-speaker datasets. Since the Ref is used as the reference to calculate the cosine similarity with the synthesized speech, we record only WER and MOS-N for Ref. As depicted in Table \ref{tab:exp-vctk}, results on the VCTK dataset, DEX-TTS outperforms the previous methods in terms of objective and subjective evaluations. Although StyleTTS shows slightly better WER in seen scenarios, the difference is marginal compared to other metrics. DEX-TTS consistently achieves high COS and MOS-S across all scenarios, indicating its ability to effectively capture and reflect rich styles from reference speech. In particular, we observe the high generalization ability of our style modeling since DEX-TTS also shows superior COS and MOS-S in zero-shot. Furthermore, the improved WER performance demonstrates that DEX-TTS can obtain enriched text representations and reflect styles without compromising text information. The outstanding MOS results suggest that DEX-TTS can synthesize reference-style speech with high fidelity. However, pooling-based single-style utilization (MetaStyleSpeech and StyleTTS) and summation- or concatenation-based style reflection methods (YourTTS and GenerSpeech) are not effective for synthesizing reference-style speech.

To verify the ability of the model to handle the styles of expressive speech, we conduct experiments on the ESD dataset in Table \ref{tab:exp-esd}. Similar to the results on the VCTK dataset, DEX-TTS outperforms previous TTS methods. It suggests that our style modeling, which handles styles based on time variability, is also effective for expressive reference speech. The outstanding performance of COS and MOS-S in the unseen scenarios indicates a strong generalization ability of DEX-TTS even in the emotional dataset. Furthermore, we observe that DEX-TTS can reflect styles without compromising speech quality compared to previous methods. Finally, unlike previous methods (YourTTS, GenerSpeech, and StyleTTS) that rely on pre-training strategies, DEX-TTS achieves excellent performance without dependence on pre-trained models. It suggests that DEX-TTS can be easily extended to various applications as a standalone model. In Section \ref{sec:more information}, we provide results with error bars for the above experiments.

%%%%%%%%%%%%%%%%%%%%%%%%%%%%%%%
\begin{table}
\caption{Ablation studies on the ESD dataset.}\label{tab:ab-esd}
\centering
\resizebox{0.65\linewidth}{!}{%
% \begin{tabular}
\begin{tabular}{p{3.8cm}M{1.2cm}M{1.2cm}M{1.2cm}M{1.2cm}}
\toprule
\multirow{2.5}{*}{Model} & \multicolumn{2}{c}{Seen} & \multicolumn{2}{c}{Unseen} \\ 
\cmidrule{2-5} & WER & COS & WER & COS  \\
\midrule
DEX-TTS               & 8.34 & 82.71 & 8.35  & 75.58   \\
\midrule
{\hspace{0pt} a) T-IV adapter $\rightarrow$ AdaIN \hspace{0pt plus 1filll}} 
                      & 11.91 & 82.75 & 10.28 & 75.29    \\
{\hspace{0pt} b) w/o T-IV styles ($h_{inv}$) \hspace{0pt plus 1filll}} 
                      & 10.84 & 82.62 & 11.05 & 74.81    \\
{\hspace{0pt} c) w/o T-V styles ($h^{e}_{v}$) \hspace{0pt plus 1filll}} 
                      & 12.19 & 78.26 & 11.72 & 71.91    \\
{\hspace{0pt} d) w/o T-V styles ($h^{d}_{v}$) \hspace{0pt plus 1filll}} 
                      & 9.51  & 82.41 & 9.18  & 74.90    \\
{\hspace{0pt} e) w/o pitch ($h_{f0}$) \hspace{0pt plus 1filll}} 
                      & 12.7 & 81.48 & 10.95 & 74.72    \\
{\hspace{0pt} f) w/o VQ \hspace{0pt plus 1filll}} 
                      & 15.70 & 82.53 & 16.84 & 76.38    \\
{\hspace{0pt} g) w/o $t$ for adapters \hspace{0pt plus 1filll}} 
                      & 9.08  & 82.13 & 9.15 & 74.65    \\
\bottomrule
\end{tabular}}

\end{table}
% %%%%%%%%%%%%%%%%%%%%%%%%%%%%%%%%

\subsection{Ablation Studies} 

To investigate the effect of the components of DEX-TTS, we conduct ablation studies in Table \ref{tab:ab-esd}. First, we analyze the effect of the T-IV adapter by replacing it with a simple AdaIN in experiment a). While the T-IV adapter utilizes all the feature maps extracted from the encoder, AdaIN only employs the last feature map. The results show considerable degradation in WER. This suggests that utilizing common features appearing in multi-level feature maps as T-IV styles is more effective. In addition, it enables to obtain well-refined styles that do not affect other speech qualities such as text content.

Experiments from b) to d) show the performance when each style, separated according to time variability, is removed. We observe that T-V and T-IV styles significantly impact WER and COS. It indicates the effectiveness of our approach which distinguishes and processes styles based on their time variability. Moreover, the most significant performance degradation is observed when removing $h^{e}_{v}$, suggesting the importance of incorporating style in the text encoder. Since the output of the text encoder is used as the initial mel representation for the prior loss calculation, style reflection in the text encoder has a considerable effect. Furthermore, the results of experiment e) show the necessity of injecting the pitch information of the reference into the T-V styles. It suggests that pitch information contains additional time-variant styles that cannot be extracted solely from the reference mel-spectrograms.

We observe interesting results for experiment f) in which the VQ layer for T-V styles $h^{d}_{v}$ is removed. Although an improvement in the COS in unseen scenarios is observed, there is a significant overall decrease in WER. To preserve the temporal information while incorporating $h^{d}_{v}$, we designed a cross-attention-based T-V adapter. However, when the VQ layer is not applied, it includes excessively detailed style information, improving similarity but significantly degrading other aspects of speech quality. Thus, the VQ layer contributes to obtaining a well-refined time-variant style, enabling an effective reflection of style information while preserving temporal details. The experiment g) demonstrates the results of removing our time step conditioning from the adapters in the diffusion decoder. Overall performance decrease is observed, highlighting the necessity of time step conditioning in adaptively incorporating styles during the iterative denoising process of the diffusion network.

\begin{table}[t]
\caption{Results on the LJSpeech test set (left) and ablation studies for encoding types on the LJSpeech test set (right). $\dagger$ indicates the overlapped patch strategy is applied.}
\centering
\begin{minipage}[t]{0.3\textwidth}
\centering
\resizebox{1.55\linewidth}{!}{%
\begin{tabular}{l|cccc}
\toprule
Model & WER & COS & MOS-N \\
\midrule
GT & 6.56 & - & 4.60 \\
FastSpeech2 \cite{ren2020fastspeech} & 7.70  & 91.31 & 3.16  \\
Grad-TTS \cite{popov2021grad} & 7.70 & 91.37 & 4.16  \\
ComoSpeech \cite{ye2023comospeech} & 8.21 & 91.58 & 4.13 \\
GeDEX-TTS$^{\dagger}$ (ours) & \textbf{6.55} & \textbf{91.75} & \textbf{4.26} \\
\bottomrule
\end{tabular}
}
\label{tab:ge-ljspeech}
\end{minipage}
\hfill
\begin{minipage}[t]{0.52\textwidth}
\centering
\resizebox{\linewidth}{!}{%
\begin{tabular}{l|l|cc}
\toprule
Model & Encoding Type & WER & COS \\
\midrule
\multirow{4}{*}{GeDEX-TTS} & sin-cos & 16.37 & 69.41  \\
& time-freq & 8.01  & 88.15  \\
& pos-freq  & 11.83 & 81.14  \\
& conv-freq & 7.31  & 91.66  \\
\midrule
GeDEX-TTS$^{\dagger}$ & conv-freq & 6.55  & 91.75  \\
\bottomrule
\end{tabular}
}
\label{tab:ge-ab-ljspeech}
\end{minipage}
\end{table}

\subsection{Further Experiments}
\label{sec:experiment-further}

As discussed in Section \ref{sec:intro}, another focus of this study is designing a strong TTS backbone. We improved diffusion-based TTS via overlapping patch strategy and conv-freq embedding, which enables the comprehensive utilization of DiT. To investigate the improvements in our diffusion network, we conduct experiments for general TTS which does not use reference speech. We eliminate the modules dependent on reference (See Section \ref{sec:appendix-GeDEX-TTS} for details), thus the model can operate as general TTS and we call this version General DEX-TTS (GeDEX-TTS). For comparison, we select previous diffusion-based TTS models and train models on a single-speaker dataset, LJSpeech \cite{ljspeech17}, following the set split of \cite{popov2021grad}. FastSpeech2 is adopted for comparison since it is a popular baseline model in general TTS. We consider 2000 epochs and $P$ of 4 for GeDEX-TTS and other training settings are the same as DEX-TTS. For inference, we use the NFE of 50 with Euler solver for all diffusion models. MOS-N is recorded by evaluations of 16 participants.

As shown in Table \ref{tab:ge-ljspeech} (left), GeDEX-TTS achieves the best performance compared to the previous methods in both objective and subjective evaluations. By leveraging patchify and embedding strategies, GeDEX-TTS effectively utilizes the structural advantages of DiT, resulting in superior performance compared to simple U-Net-based diffusion models (i.e., Grad-TTS and CoMoSpeech). Notably, the WER performance of GeDEX-TTS is on par with that of the Ground Truth (GT), showing the validity of network improvement in diffusion TTS. The results reveal that improvements in the diffusion network are consistently effective beyond expressive TTS to general TTS as well, indicating that the proposed method also exhibits considerable significance as a general TTS network.

To analyze the effect of the network improvement strategies, we conduct ablation studies on the LJSpeech dataset. The first block of Table \ref{tab:ge-ab-ljspeech} (right) shows the results depending on different ways of encoding for patch embedding. Instead of conv-freq embedding used in our model, we apply other popular methods: \textbf{1) sin-cos}, frequency-based positional
embeddings \cite{dosovitskiy2020image}. \textbf{2) time-freq}, fixed size learnable parameters for time and frequency axis \cite{koutini2021efficient}. \textbf{3) pos-freq}, positional encoding for the time axis and fixed size learnable parameters for the frequency axis (we added it to compare with conv-freq). The comparison results show lower performance of the conventional encoding types despite their stable performance when using fixed image sizes in image synthesis. This suggests that relative patch embedding using convolution is more suitable for tasks with significant variations in the temporal axis length, such as speech synthesis. Lastly, the results of the second block suggest that the overlapping patchify strategy contributes to synthesizing more natural speech by mitigating the boundary artifacts between patches. 

% These strategies enable more effective utilization of DiT.

\section{Conclusion}

In this study, we proposed DEX-TTS, a reference-based TTS, which can synthesize high-quality and reference-style speech. First, we improved the diffusion-based TTS backbone by overlapping patchify and conv-freq embedding strategies, which enable the effective utilization of DiT architecture. To extract well-represented styles from the reference, we categorized the styles into time-invariant and time-variant styles, with T-IV and T-V encoders using multi-level feature maps and vector quantization for obtaining well-refined styles in each manner. We designed adapters with adaptive normalization and cross-attention methods for effective style reflection with high generalization ability. The experimental results on the VCTK and ESD datasets suggest that DEX-TTS, even without using pre-trained strategies, outperformed the previous expressive TTS models. In addition, DEX-TTS consistently exhibited superior performance across all metrics, indicating its effective style reflection ability which did not compromise speech quality, unlike other models. Lastly, to validate our strategies for improving the diffusion network, we conducted experiments using our diffusion TTS backbone in a general TTS task. The results on the LJSpeech dataset demonstrated that our diffusion backbone also achieved outstanding performance in the general TTS task.

\begin{ack}
This research was supported by a Korea TechnoComplex Foundation Grant (R2112653) and Korea University Grant (K2403371). This research was also supported by Brain Korea 21 FOUR.
\end{ack}

%% ** possible option: [unsrt, siam, plain, ieeetr, apalike, alpha, acm, abbrv]
{
\small
\bibliographystyle{unsrt}
\bibliography{paper}

\begin{thebibliography}{10}

\bibitem{oord2016wavenet}
Aaron van~den Oord, Sander Dieleman, Heiga Zen, Karen Simonyan, Oriol Vinyals, Alex Graves, Nal Kalchbrenner, Andrew Senior, and Koray Kavukcuoglu.
\newblock Wavenet: A generative model for raw audio.
\newblock {\em arXiv preprint arXiv:1609.03499}, 2016.

\bibitem{wang2017tacotron}
Yuxuan Wang, RJ~Skerry-Ryan, Daisy Stanton, Yonghui Wu, Ron~J Weiss, Navdeep Jaitly, Zongheng Yang, Ying Xiao, Zhifeng Chen, Samy Bengio, et~al.
\newblock Tacotron: Towards end-to-end speech synthesis.
\newblock {\em arXiv preprint arXiv:1703.10135}, 2017.

\bibitem{shen2018natural}
Jonathan Shen, Ruoming Pang, Ron~J Weiss, Mike Schuster, Navdeep Jaitly, Zongheng Yang, Zhifeng Chen, Yu~Zhang, Yuxuan Wang, Rj~Skerrv-Ryan, et~al.
\newblock Natural tts synthesis by conditioning wavenet on mel spectrogram predictions.
\newblock In {\em 2018 IEEE international conference on acoustics, speech and signal processing (ICASSP)}, pages 4779--4783. IEEE, 2018.

\bibitem{ren2019fastspeech}
Yi~Ren, Yangjun Ruan, Xu~Tan, Tao Qin, Sheng Zhao, Zhou Zhao, and Tie-Yan Liu.
\newblock Fastspeech: Fast, robust and controllable text to speech.
\newblock {\em Advances in neural information processing systems}, 32, 2019.

\bibitem{li2019neural}
Naihan Li, Shujie Liu, Yanqing Liu, Sheng Zhao, and Ming Liu.
\newblock Neural speech synthesis with transformer network.
\newblock In {\em Proceedings of the AAAI conference on artificial intelligence}, 2019.

\bibitem{yamamoto2020parallel}
Ryuichi Yamamoto, Eunwoo Song, and Jae-Min Kim.
\newblock Parallel wavegan: A fast waveform generation model based on generative adversarial networks with multi-resolution spectrogram.
\newblock In {\em ICASSP 2020-2020 IEEE International Conference on Acoustics, Speech and Signal Processing (ICASSP)}, pages 6199--6203. IEEE, 2020.

\bibitem{yang2021ganspeech}
Jinhyeok Yang, Jae-Sung Bae, Taejun Bak, Youngik Kim, and Hoon-Young Cho.
\newblock Ganspeech: Adversarial training for high-fidelity multi-speaker speech synthesis.
\newblock {\em arXiv preprint arXiv:2106.15153}, 2021.

\bibitem{miao2020flow}
Chenfeng Miao, Shuang Liang, Minchuan Chen, Jun Ma, Shaojun Wang, and Jing Xiao.
\newblock Flow-tts: A non-autoregressive network for text to speech based on flow.
\newblock In {\em ICASSP 2020-2020 IEEE International Conference on Acoustics, Speech and Signal Processing (ICASSP)}, pages 7209--7213. IEEE, 2020.

\bibitem{kim2020glow}
Jaehyeon Kim, Sungwon Kim, Jungil Kong, and Sungroh Yoon.
\newblock Glow-tts: A generative flow for text-to-speech via monotonic alignment search.
\newblock {\em Advances in Neural Information Processing Systems}, 33:8067--8077, 2020.

\bibitem{ho2020denoising}
Jonathan Ho, Ajay Jain, and Pieter Abbeel.
\newblock Denoising diffusion probabilistic models.
\newblock {\em Advances in neural information processing systems}, 33:6840--6851, 2020.

\bibitem{ho2022imagen}
Jonathan Ho, William Chan, Chitwan Saharia, Jay Whang, Ruiqi Gao, Alexey Gritsenko, Diederik~P Kingma, Ben Poole, Mohammad Norouzi, David~J Fleet, et~al.
\newblock Imagen video: High definition video generation with diffusion models.
\newblock {\em arXiv preprint arXiv:2210.02303}, 2022.

\bibitem{schneider2023archisound}
Flavio Schneider.
\newblock Archisound: Audio generation with diffusion.
\newblock {\em arXiv preprint arXiv:2301.13267}, 2023.

\bibitem{kong2020diffwave}
Zhifeng Kong, Wei Ping, Jiaji Huang, Kexin Zhao, and Bryan Catanzaro.
\newblock Diffwave: A versatile diffusion model for audio synthesis.
\newblock {\em arXiv preprint arXiv:2009.09761}, 2020.

\bibitem{jeong2021diff}
Myeonghun Jeong, Hyeongju Kim, Sung~Jun Cheon, Byoung~Jin Choi, and Nam~Soo Kim.
\newblock Diff-tts: A denoising diffusion model for text-to-speech.
\newblock {\em arXiv preprint arXiv:2104.01409}, 2021.

\bibitem{popov2021grad}
Vadim Popov, Ivan Vovk, Vladimir Gogoryan, Tasnima Sadekova, and Mikhail Kudinov.
\newblock Grad-tts: A diffusion probabilistic model for text-to-speech.
\newblock In {\em International Conference on Machine Learning}, pages 8599--8608. PMLR, 2021.

\bibitem{liu2022diffgan}
Songxiang Liu, Dan Su, and Dong Yu.
\newblock Diffgan-tts: High-fidelity and efficient text-to-speech with denoising diffusion gans.
\newblock {\em arXiv preprint arXiv:2201.11972}, 2022.

\bibitem{tan2021survey}
Xu~Tan, Tao Qin, Frank Soong, and Tie-Yan Liu.
\newblock A survey on neural speech synthesis.
\newblock {\em arXiv preprint arXiv:2106.15561}, 2021.

\bibitem{lee2017emotional}
Younggun Lee, Azam Rabiee, and Soo-Young Lee.
\newblock Emotional end-to-end neural speech synthesizer.
\newblock {\em arXiv preprint arXiv:1711.05447}, 2017.

\bibitem{kim2021expressive}
Minchan Kim, Sung~Jun Cheon, Byoung~Jin Choi, Jong~Jin Kim, and Nam~Soo Kim.
\newblock Expressive text-to-speech using style tag.
\newblock {\em arXiv preprint arXiv:2104.00436}, 2021.

\bibitem{li2021controllable}
Tao Li, Shan Yang, Liumeng Xue, and Lei Xie.
\newblock Controllable emotion transfer for end-to-end speech synthesis.
\newblock In {\em 2021 12th International Symposium on Chinese Spoken Language Processing (ISCSLP)}, pages 1--5. IEEE, 2021.

\bibitem{wang2018style}
Yuxuan Wang, Daisy Stanton, Yu~Zhang, RJ-Skerry Ryan, Eric Battenberg, Joel Shor, Ying Xiao, Ye~Jia, Fei Ren, and Rif~A Saurous.
\newblock Style tokens: Unsupervised style modeling, control and transfer in end-to-end speech synthesis.
\newblock In {\em International conference on machine learning}, pages 5180--5189. PMLR, 2018.

\bibitem{valle2020mellotron}
Rafael Valle, Jason Li, Ryan Prenger, and Bryan Catanzaro.
\newblock Mellotron: Multispeaker expressive voice synthesis by conditioning on rhythm, pitch and global style tokens.
\newblock In {\em ICASSP 2020-2020 IEEE International Conference on Acoustics, Speech and Signal Processing (ICASSP)}, pages 6189--6193. IEEE, 2020.

\bibitem{lee2021styler}
Keon Lee, Kyumin Park, and Daeyoung Kim.
\newblock Styler: Style factor modeling with rapidity and robustness via speech decomposition for expressive and controllable neural text to speech.
\newblock {\em arXiv preprint arXiv:2103.09474}, 2021.

\bibitem{min2021meta}
Dongchan Min, Dong~Bok Lee, Eunho Yang, and Sung~Ju Hwang.
\newblock Meta-stylespeech: Multi-speaker adaptive text-to-speech generation.
\newblock In {\em International Conference on Machine Learning}, pages 7748--7759. PMLR, 2021.

\bibitem{casanova2022yourtts}
Edresson Casanova, Julian Weber, Christopher~D Shulby, Arnaldo~Candido Junior, Eren G{\"o}lge, and Moacir~A Ponti.
\newblock Yourtts: Towards zero-shot multi-speaker tts and zero-shot voice conversion for everyone.
\newblock In {\em International Conference on Machine Learning}, pages 2709--2720. PMLR, 2022.

\bibitem{li2022styletts}
Yinghao~Aaron Li, Cong Han, and Nima Mesgarani.
\newblock Styletts: A style-based generative model for natural and diverse text-to-speech synthesis.
\newblock {\em arXiv preprint arXiv:2205.15439}, 2022.

\bibitem{huang2022generspeech}
Rongjie Huang, Yi~Ren, Jinglin Liu, Chenye Cui, and Zhou Zhao.
\newblock Generspeech: Towards style transfer for generalizable out-of-domain text-to-speech synthesis.
\newblock {\em arXiv preprint arXiv:2205.07211}, 2022.

\bibitem{ye2023comospeech}
Zhen Ye, Wei Xue, Xu~Tan, Jie Chen, Qifeng Liu, and Yike Guo.
\newblock Comospeech: One-step speech and singing voice synthesis via consistency model.
\newblock {\em arXiv preprint arXiv:2305.06908}, 2023.

\bibitem{jing2023u}
Xin Jing, Yi~Chang, Zijiang Yang, Jiangjian Xie, Andreas Triantafyllopoulos, and Bjoern~W Schuller.
\newblock U-dit tts: U-diffusion vision transformer for text-to-speech.
\newblock {\em arXiv preprint arXiv:2305.13195}, 2023.

\bibitem{peebles2023scalable}
William Peebles and Saining Xie.
\newblock Scalable diffusion models with transformers.
\newblock In {\em Proceedings of the IEEE/CVF International Conference on Computer Vision}, pages 4195--4205, 2023.

\bibitem{huang2017arbitrary}
Xun Huang and Serge Belongie.
\newblock Arbitrary style transfer in real-time with adaptive instance normalization.
\newblock In {\em Proceedings of the IEEE international conference on computer vision}, pages 1501--1510, 2017.

\bibitem{vaswani2017attention}
Ashish Vaswani, Noam Shazeer, Niki Parmar, Jakob Uszkoreit, Llion Jones, Aidan~N Gomez, {\L}ukasz Kaiser, and Illia Polosukhin.
\newblock Attention is all you need.
\newblock {\em Advances in neural information processing systems}, 30, 2017.

\bibitem{song2020score}
Yang Song, Jascha Sohl-Dickstein, Diederik~P Kingma, Abhishek Kumar, Stefano Ermon, and Ben Poole.
\newblock Score-based generative modeling through stochastic differential equations.
\newblock {\em arXiv preprint arXiv:2011.13456}, 2020.

\bibitem{karras2022elucidating}
Tero Karras, Miika Aittala, Timo Aila, and Samuli Laine.
\newblock Elucidating the design space of diffusion-based generative models.
\newblock {\em Advances in Neural Information Processing Systems}, 35:26565--26577, 2022.

\bibitem{su2024roformer}
Jianlin Su, Murtadha Ahmed, Yu~Lu, Shengfeng Pan, Wen Bo, and Yunfeng Liu.
\newblock Roformer: Enhanced transformer with rotary position embedding.
\newblock {\em Neurocomputing}, 568:127063, 2024.

\bibitem{sun2023retentive}
Yutao Sun, Li~Dong, Shaohan Huang, Shuming Ma, Yuqing Xia, Jilong Xue, Jianyong Wang, and Furu Wei.
\newblock Retentive network: A successor to transformer for large language models.
\newblock {\em arXiv preprint arXiv:2307.08621}, 2023.

\bibitem{chen2021adaspeech}
Mingjian Chen, Xu~Tan, Bohan Li, Yanqing Liu, Tao Qin, Sheng Zhao, and Tie-Yan Liu.
\newblock Adaspeech: Adaptive text to speech for custom voice.
\newblock {\em arXiv preprint arXiv:2103.00993}, 2021.

\bibitem{park2023triaan}
Hyun~Joon Park, Seok~Woo Yang, Jin~Sob Kim, Wooseok Shin, and Sung~Won Han.
\newblock Triaan-vc: Triple adaptive attention normalization for any-to-any voice conversion.
\newblock In {\em ICASSP 2023-2023 IEEE International Conference on Acoustics, Speech and Signal Processing (ICASSP)}, pages 1--5. IEEE, 2023.

\bibitem{van2017neural}
Aaron Van Den~Oord, Oriol Vinyals, et~al.
\newblock Neural discrete representation learning.
\newblock {\em Advances in neural information processing systems}, 30, 2017.

\bibitem{vctk2019}
Junichi Yamagishi, Christophe Veaux, and Kirsten MacDonald.
\newblock Cstr vctk corpus: English multi-speaker corpus for cstr voice cloning toolkit (version 0.92).
\newblock In {\em University of Edinburgh. The Centre for Speech Technology Research (CSTR)}, 2019.

\bibitem{zhou2021seen}
Kun Zhou, Berrak Sisman, Rui Liu, and Haizhou Li.
\newblock Seen and unseen emotional style transfer for voice conversion with a new emotional speech dataset.
\newblock In {\em ICASSP 2021-2021 IEEE International Conference on Acoustics, Speech and Signal Processing (ICASSP)}, pages 920--924. IEEE, 2021.

\bibitem{kong2020hifi}
Jungil Kong, Jaehyeon Kim, and Jaekyoung Bae.
\newblock Hifi-gan: Generative adversarial networks for efficient and high fidelity speech synthesis.
\newblock {\em Advances in Neural Information Processing Systems}, 33:17022--17033, 2020.

\bibitem{baevski2020wav2vec}
Alexei Baevski, Yuhao Zhou, Abdelrahman Mohamed, and Michael Auli.
\newblock wav2vec 2.0: A framework for self-supervised learning of speech representations.
\newblock {\em Advances in neural information processing systems}, 33:12449--12460, 2020.

\bibitem{ren2020fastspeech}
Yi~Ren, Chenxu Hu, Xu~Tan, Tao Qin, Sheng Zhao, Zhou Zhao, and Tie-Yan Liu.
\newblock Fastspeech 2: Fast and high-quality end-to-end text to speech.
\newblock {\em arXiv preprint arXiv:2006.04558}, 2020.

\bibitem{ljspeech17}
Keith Ito and Linda Johnson.
\newblock The lj speech dataset.
\newblock \url{https://keithito.com/LJ-Speech-Dataset/}, 2017.

\bibitem{dosovitskiy2020image}
Alexey Dosovitskiy, Lucas Beyer, Alexander Kolesnikov, Dirk Weissenborn, Xiaohua Zhai, Thomas Unterthiner, Mostafa Dehghani, Matthias Minderer, Georg Heigold, Sylvain Gelly, et~al.
\newblock An image is worth 16x16 words: Transformers for image recognition at scale.
\newblock {\em arXiv preprint arXiv:2010.11929}, 2020.

\bibitem{koutini2021efficient}
Khaled Koutini, Jan Schl{\"u}ter, Hamid Eghbal-Zadeh, and Gerhard Widmer.
\newblock Efficient training of audio transformers with patchout.
\newblock {\em arXiv preprint arXiv:2110.05069}, 2021.

\bibitem{song2023consistency}
Yang Song, Prafulla Dhariwal, Mark Chen, and Ilya Sutskever.
\newblock Consistency models.
\newblock {\em arXiv preprint arXiv:2303.01469}, 2023.

\bibitem{kim2023consistency}
Dongjun Kim, Chieh-Hsin Lai, Wei-Hsiang Liao, Naoki Murata, Yuhta Takida, Toshimitsu Uesaka, Yutong He, Yuki Mitsufuji, and Stefano Ermon.
\newblock Consistency trajectory models: Learning probability flow ode trajectory of diffusion.
\newblock {\em arXiv preprint arXiv:2310.02279}, 2023.

\end{thebibliography}
}

%%%%%%%%%%%%%%%%%%%%%%%%%%%%%%%%%%%%%%%%%%%%%%%%%%%%%%%%%%%%

\newpage
\appendix
\section*{\Large Appendix / supplemental material}
\section{Details of DEX-TTS}

In this section, we provide further information about DEX-TTS. Specifically, we describe in detail the text encoder process and GeDEX-TTS.

\begin{figure*}[ht]
  \centering
  \includegraphics[scale=0.65]{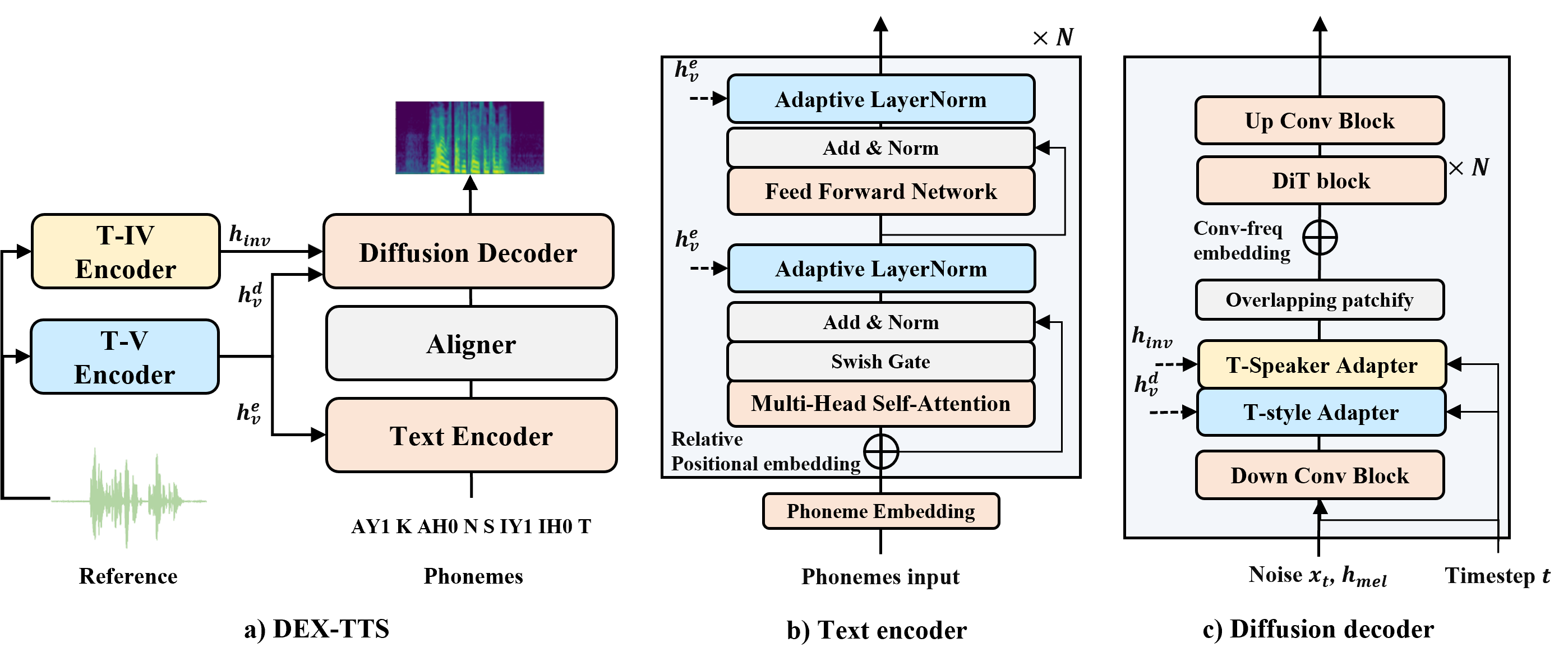} 
  \caption{Overall architecture of DEX-TTS, Text encoder, and Diffusion decoder.} 
  \label{fig:dextts2}
\end{figure*}

\subsection{Text Encoder}
\label{sec:appendix-text encoder}

As depicted in Figure \ref{fig:dextts2}, the text encoder of DEX-TTS consists of $N$ Transformer encoder layers. We incorporate relative position embedding, RoPE \cite{su2024roformer}, into the attention mechanism, and apply the swish gate used in RetNet \cite{sun2023retentive} after the attention operation to improve the text encoder. In addition, AdaLN \cite{chen2021adaspeech} is used to condition time-variant (T-V) style $h^{e}_{v}$.

Let $X \in \mathbb{R}^{L_{p} \times C}$ be the initial text representation from phonemes embedding, where $L_{p}$ is the phonemes lengths and $C$ is the hidden size. Before describing Multi-Head Self-Attention (MHSA), our self-attention mechanism with RoPE, which injects the absolute position encoding by rotations and keeps relative position by the inner product, is defined as follows:
%###########################
\begin{equation}\label{eq:sa}
  \begin{gathered}
    Q=XW_{q}\Theta, \ \ K=XW_{k}\bar{\Theta}, \ \ V = XW_{v}, \ \ \Theta_{n}=e^{in\theta} \\ 
    Attention(Q,K,V)=softmax(QK^{\top} / \sqrt{d_{k}})V
  \end{gathered}
\end{equation}
%##########################
where $W_{q,k,v} \in \mathbb{R}^{C \times C}$ is the linear weight for the projection, and $\sqrt{d_{k}}$ is the scaling factor. $n$ is the absolute position number in lengths $L_{p}$, and $\bar{\Theta}$ is the complex conjugate of $\Theta$. Given $h$ is the number of attention heads, we extend Equation \ref{eq:sa} to MHSA with a swish gate as follows:
%###########################
\begin{equation}\label{eq:mhsa}
  \begin{gathered}
    head_{i}=Attention(Q_{i},K_{i},V_{i}) \\
    Y = GN_{h}([head_{1};,...,;head_{h}]) \\
    MHSA(X) = swish(XW_{g})YW_{o}
  \end{gathered}
\end{equation}
%##########################
$Q_{i}$, $K_{i}$, and $V_{i}$ are used to compute each attention head, each having dimensions divided from the original dimension by the number of heads. Then, Group Normalization ($GN$) is applied to the concatenated heads. Lastly, we utilize a swish gate, where $W_{g,o} \in \mathbb{R}^{C \times C}$ is the linear weight for projection. Based on the MHSA, the text encoder layer processes with residual connection, layer normalization ($LN$), and AdaLN are defined in Equation \ref{eq:encoder layer}.
%###########################
\begin{equation}\label{eq:encoder layer}
  \begin{gathered}
    Y = MHSA(LN(X))+X \\
    Y = AdaLN(Y, h^{e}_{v}) \\
    X' = FFN(LN(Y))+Y \\
    X' = AdaLN(X', h^{e}_{v})
  \end{gathered}
\end{equation}
%##########################
where $h^{e}_{v}$ is T-V style from the T-V encoder, and $FFN$ is the feed-forward network which consists of two linear weights with GeLU activation. In the experiments, we use $N$ of 8, $C$ of 192, and $h$ of 2.

\begin{figure*}[ht]
  \centering
  \includegraphics[scale=0.65]{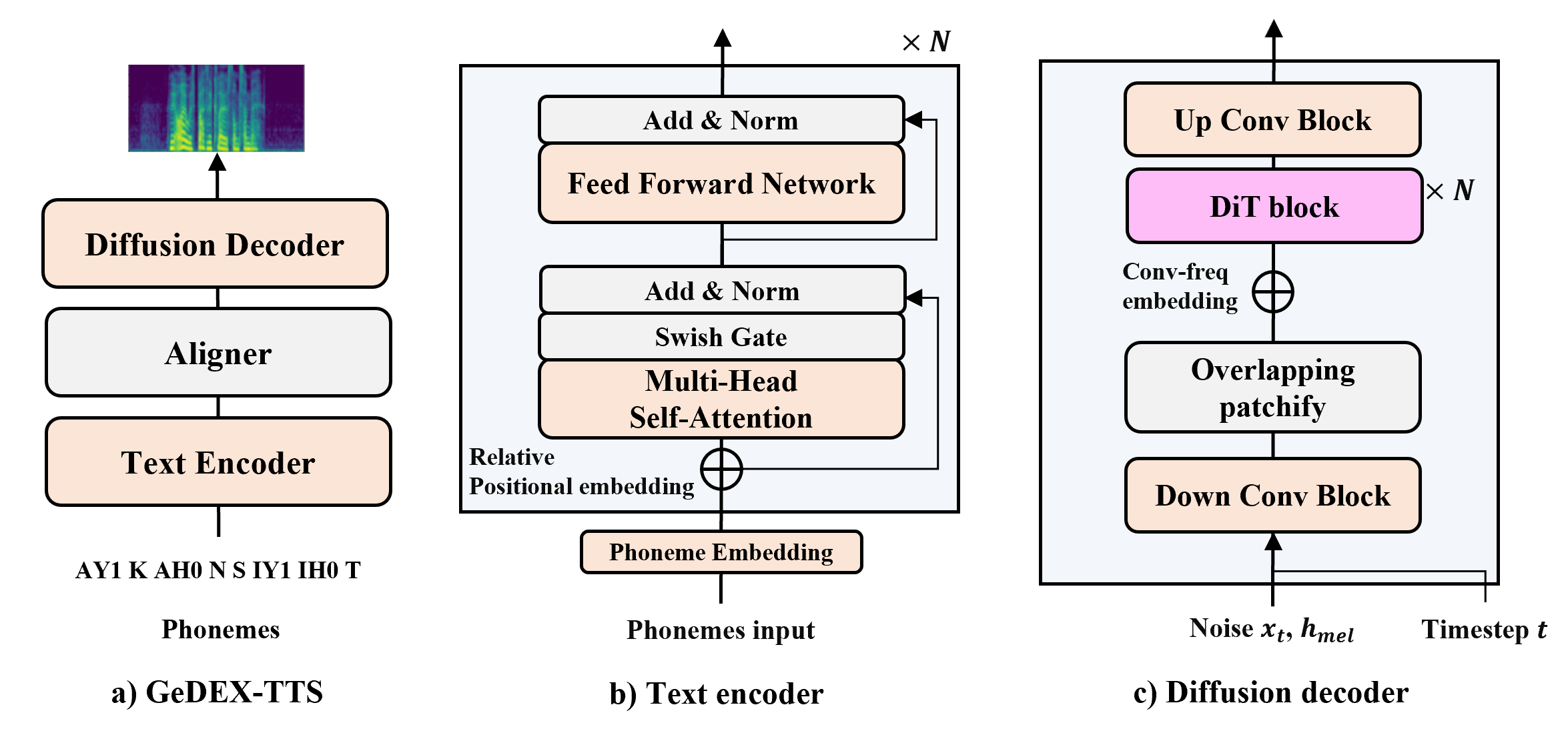} 
  \caption{Overall architecture of GeDEX-TTS.} 
  \label{fig:gedextts}
\end{figure*}

\subsection{GeDEX-TTS}
\label{sec:appendix-GeDEX-TTS}

To verify improvements in diffusion-based TTS backbone, we designed General DEX-TTS (GeDEX-TTS) which synthesizes speech without references, and we conducted experiments in Section \ref{sec:experiment-further}. To enable GeDEX-TTS to operate without a reference, we removed modules dependent on the reference. As shown in Figure \ref{fig:gedextts}, we removed T-IV and T-V encoders. In addition, AdaLN in the text encoder and each adapter in the diffusion decoder are removed. With the exception of the removed modules, the other components are the same as DEX-TTS.

\section{Additional Analysis of Experimental Results}

%%%%%%%%%%%%%%%%%%%%%%%%%%%%%%%
\begin{table}[ht]
\caption{Comparison results on the VCTK dataset (Table \ref{tab:exp-vctk}) with error ranges.}\label{tab:exp-vctk-appendix}
\centering
\resizebox{0.9\columnwidth}{!}{%
\begin{tabular}{p{3.5cm}M{1.2cm}M{1.2cm}M{1.2cm}M{1.2cm}cM{1.2cm}M{1.2cm}M{1.2cm}M{1.2cm}}
% \begin{tabular}{p{2.5cm}ccccccccccc}
\toprule
\multirow{2.5}{*}{Model} & \multicolumn{4}{c}{Seen scenarios} & & \multicolumn{4}{c}{Unseen (zero-shot) scenarios} \\ 
\cmidrule{2-5} \cmidrule{7-10} & WER & COS & MOS-N & MOS-S & & WER & COS & MOS-N & MOS-S \\
\midrule

Ref                                      & 6.23  \newline $\pm$ 0.58 & -      
                                         & 3.97  \newline $\pm$ 0.04 & - & 
                                         & 6.23  \newline $\pm$ 0.58 & -      
                                         & 3.97  \newline $\pm$ 0.04 & -  \\
MetaStyleSpeech \cite{min2021meta}       & 16.58 \newline $\pm$ 1.20 & 78.10 \newline $\pm$ 0.39 
                                         & 3.43  \newline $\pm$ 0.05 & 3.63  \newline $\pm$ 0.06 & 
                                         & 16.5  \newline $\pm$ 1.18 & 73.53 \newline $\pm$ 0.40 
                                         & 3.38  \newline $\pm$ 0.06 & 3.30  \newline $\pm$ 0.06  \\
YourTTS \cite{casanova2022yourtts}       & 21.27 \newline $\pm$ 1.26 & 78.78 \newline $\pm$ 0.28
                                         & 3.20  \newline $\pm$ 0.05 & 3.11  \newline $\pm$ 0.06 & 
                                         & 18.34 \newline $\pm$ 1.28 & 75.00 \newline $\pm$ 0.31
                                         & 3.33  \newline $\pm$ 0.05 & 3.08  \newline $\pm$ 0.06  \\
GenerSpeech \cite{huang2022generspeech}  & 13.87 \newline $\pm$ 1.21 & 77.46 \newline $\pm$ 0.41
                                         & 3.40  \newline $\pm$ 0.05 & 3.25  \newline $\pm$ 0.05 & 
                                         & 11.37 \newline $\pm$ 1.09 & 73.23 \newline $\pm$ 0.55
                                         & 3.46  \newline $\pm$ 0.05 & 3.06  \newline $\pm$ 0.06  \\
StyleTTS \cite{li2022styletts}           & \textbf{7.72} \newline $\pm$ 0.72  & 82.93 \newline $\pm$ 0.26
                                         & 3.57  \newline $\pm$ 0.05 & 3.70  \newline $\pm$ 0.06 & 
                                         & 6.58  \newline $\pm$ 0.72 & 77.90 \newline $\pm$ 0.31
                                         & 3.53  \newline $\pm$ 0.05 & 3.65  \newline $\pm$ 0.06  \\

\midrule
DEX-TTS                                  & 7.85 \newline $\pm$ 0.73 & \textbf{85.31} \newline $\pm$ 0.22
                                         & \textbf{3.75} \newline $\pm$ 0.05 & \textbf{3.88}  \newline $\pm$ 0.05 & 
                                         & \textbf{5.84} \newline $\pm$ 0.61 & \textbf{80.45} \newline $\pm$ 0.26
                                         & \textbf{3.76} \newline $\pm$ 0.04 & \textbf{3.81}  \newline $\pm$ 0.06 \\
\bottomrule
\end{tabular}}
\end{table}
% %%%%%%%%%%%%%%%%%%%%%%%%%%%%%%%%

%%%%%%%%%%%%%%%%%%%%%%%%%%%%%%%
\begin{table}[ht]
\caption{Comparison results on the ESD dataset (Table \ref{tab:exp-esd}) with error ranges.}\label{tab:exp-esd-appendix}
\centering
\resizebox{0.9\columnwidth}{!}{%
\begin{tabular}{p{3.5cm}M{1.2cm}M{1.2cm}M{1.2cm}M{1.2cm}cM{1.2cm}M{1.2cm}M{1.2cm}M{1.2cm}}
% \begin{tabular}{p{2.5cm}ccccccccccc}
\toprule
\multirow{2.5}{*}{Model} & \multicolumn{4}{c}{Seen scenarios} & & \multicolumn{4}{c}{Unseen (zero-shot) scenarios} \\ 
\cmidrule{2-5} \cmidrule{7-10} & WER  & COS & MOS-N & MOS-S & & WER & COS & MOS-N & MOS-S \\
\midrule

Ref                                      & 7.12  \newline $\pm$ 0.68 & -      
                                         & 3.90  \newline $\pm$ 0.05 & - & 
                                         & 7.12  \newline $\pm$ 0.68 & -      
                                         & 3.90  \newline $\pm$ 0.05 & -  \\
MetaStyleSpeech \cite{min2021meta}       & 24.56 \newline $\pm$ 1.71 & 79.41 \newline $\pm$ 0.39  
                                         & 3.09  \newline $\pm$ 0.05 & 3.53  \newline $\pm$ 0.05 & 
                                         & 25.84 \newline $\pm$ 1.73 & 73.01 \newline $\pm$ 0.38
                                         & 3.19  \newline $\pm$ 0.05 & 3.34  \newline $\pm$ 0.05  \\
YourTTS \cite{casanova2022yourtts}       & 16.57 \newline $\pm$ 1.32 & 77.61 \newline $\pm$ 0.36
                                         & 3.33  \newline $\pm$ 0.05 & 3.40  \newline $\pm$ 0.05 & 
                                         & 16.35 \newline $\pm$ 1.25 & 69.38 \newline $\pm$ 0.37
                                         & 3.28  \newline $\pm$ 0.05 & 2.96  \newline $\pm$ 0.05  \\
GenerSpeech \cite{huang2022generspeech}  & 12.75 \newline $\pm$ 1.26 & 75.09 \newline $\pm$ 0.57
                                         & 3.23  \newline $\pm$ 0.05 & 3.28  \newline $\pm$ 0.05 & 
                                         & 11.78 \newline $\pm$ 1.1  & 70.54 \newline $\pm$ 0.47
                                         & 3.06  \newline $\pm$ 0.05 & 2.78  \newline $\pm$ 0.06  \\
StyleTTS \cite{li2022styletts}           & 12.59 \newline $\pm$ 1.08 & 79.65 \newline $\pm$ 0.34
                                         & 3.41  \newline $\pm$ 0.05 & 3.50  \newline $\pm$ 0.05 & 
                                         & 12.11 \newline $\pm$ 1.17 & 72.27 \newline $\pm$ 0.36
                                         & 3.23  \newline $\pm$ 0.05 & 3.06  \newline $\pm$ 0.05  \\
\midrule
DEX-TTS                                  & \textbf{8.34} \newline $\pm$ 0.95 & \textbf{82.71} \newline $\pm$ 0.30
                                         & \textbf{3.73} \newline $\pm$ 0.05 & \textbf{3.84}  \newline $\pm$ 0.05 & 
                                         & \textbf{8.35} \newline $\pm$ 0.93 & \textbf{75.58} \newline $\pm$ 0.31 
                                         & \textbf{3.57} \newline $\pm$ 0.05 & \textbf{3.52}  \newline $\pm$ 0.05 \\
                                         
\bottomrule
\end{tabular}}
\end{table}
% %%%%%%%%%%%%%%%%%%%%%%%%%%%%%%%%

% %%%%%%%%%%%%%%%%%%%%%%%%%%%%%%%%
\begin{table}[ht]
\caption{Results on the LJSpeech test set (Table \ref{tab:ge-ljspeech}) with error ranges.}
\centering
\resizebox{0.55\linewidth}{!}{%
\begin{tabular}{l|cccc}
\toprule
Model & WER & COS & MOS-N \\
\midrule
{GT \hspace{0pt plus 1filll}} 
    & 6.56 \ \newline $\pm$ 0.55  & - & 4.60 $\pm$ 0.04 \\
{FastSpeech2 \hspace{0pt plus 1filll}} 
    & 7.70  \ \newline $\pm$ 0.57 & 91.31  \ \newline $\pm$ 0.15 & 3.16 $\pm$ 0.06  \\
{Grad-TTS    \hspace{0pt plus 1filll}} 
    & 7.70  \ \newline $\pm$ 0.59 & 91.37  \ \newline $\pm$ 0.18 & 4.16 $\pm$ 0.05  \\
% {UDiT-TTS    \hspace{0pt plus 1filll}} 
%     & 8.91  \ \newline $\pm$ 0.61 & 86.76  \ \newline $\pm$ 0.20 & 3.22 $\pm$ 0.06  \\
{ComoSpeech  \hspace{0pt plus 1filll}} 
    & 8.21  \ \newline $\pm$ 0.60 & 91.58  \ \newline $\pm$ 0.17 & 4.13 $\pm$ 0.05 \\
{GeDEX-TTS \hspace{0pt plus 1filll}} 
    & \textbf{6.55} \ \newline $\pm$ 0.55 & \textbf{91.75} \ \newline $\pm$ 0.16 & \textbf{4.26} $\pm$ 0.05 \\
\bottomrule
\end{tabular}
}
\label{tab:ge-ljspeech-appendix}
\end{table}
% %%%%%%%%%%%%%%%%%%%%%%%%%%%%%%%%

\subsection{More Information about Experimental Results}
\label{sec:more information}

To analyze the experimental results in the main text beyond the single-dimensional summaries of performance, we further present the sample errors of each evaluation metric. We provide the information for the main experimental results of expressive TTS (on the VCTK and ESD dataset - Table \ref{tab:exp-vctk} and Table \ref{tab:exp-esd}) and general TTS (on the LJSpeech dataset - Table \ref{tab:ge-ljspeech}) conducted in the main text.

As evident from the information in Tables \ref{tab:exp-vctk-appendix}, \ref{tab:exp-esd-appendix}, and \ref{tab:ge-ljspeech-appendix}, the sample errors of the proposed models are generally lower, indicating higher stability than the previous methods. Furthermore, considering the magnitudes of the values, the performance improvements in the main text are sufficiently significant.

 % Diffusion & UDiT-TTS    & 0.26 & 52.88M \\
% %%%%%%%%%%%%%%%%%%%%%%%%%%%%%%%%
\begin{table}[ht]
\caption{Model complexities. We record the RTFs for each model including the vocoder (HiFi-GAN) process except for YourTTS which does not require the vocoder. The models requiring the vocoder additionally need the number of parameters for the vocoder (about 14M).}
\centering
\resizebox{\linewidth}{!}{%
\begin{tabular}{lll|ccc|lll|cc}
\toprule
Task & Type &  Model & RTF & \# Params & & Task & Type & Model & RTF & \# Params \\
\midrule
\multirow{5}{*}{Expressive TTS}& Non-diffusion & MetaStyleSpeech  & 0.034  & 27.67M &  & \multirow{5}{*}{General TTS}& Non-diffusion & FastSpeech2 & 0.021 & 34.65M \\
                               & Non-diffusion & YourTTS          & 0.062  & 94.60M &  &                              & Diffusion & Grad-TTS    & 0.171 & 14.84M \\
                               & Non-diffusion & GenerSpeech      & 0.138  & 51.64M &  &                             & 
                               Diffusion & ComoSpeech  & 0.178 & 14.84M \\
                               & Non-diffusion & StyleTTS         & 0.038  & 68.34M &  &                             & Diffusion & GeDEX-TTS   & 0.163 & 15.04M \\
                               & Diffusion & DEX-TTS          & 0.297  & 18.36M &  &                             & 
                               & & & \\
\bottomrule
\end{tabular}
}
\label{tab:model complexity}
\end{table}
% %%%%%%%%%%%%%%%%%%%%%%%%%%%%%%%%

% %%%%%%%%%%%%%%%%%%%%%%%%%%%%%%%%
\begin{table}[ht]
\caption{Experiment results on the VCTK and LJSpeech datasets depending on NFE. For the performance of the DEX-TTS, we take the average of seen and unseen scenarios.}
\centering
\resizebox{0.6\linewidth}{!}{%
\begin{tabular}{l|l|cccc}
\toprule
Model & NFE & WER & COS & CMOS-N & RTF  \\ 
\midrule

\multirow{3}{*}{DEX-TTS}    & 10 & 6.72 \newline $\pm$ 0.68 & 82.77 \newline $\pm$ 0.25 & -0.07 \newline $\pm$ 0.04 & 0.087  \\
                            & 25 & 7.04 \newline $\pm$ 0.70 & 82.84 \newline $\pm$ 0.24 & -0.06 \newline $\pm$ 0.04 & 0.167  \\
                            & 50 & 6.84 \newline $\pm$ 0.67 & 82.88 \newline $\pm$ 0.24 & 0 & 0.297  \\

\midrule

\multirow{3}{*}{GeDEX-TTS}  & 10  & 6.50 \newline $\pm$ 0.56 & 91.68  \newline $\pm$ 0.15 & -0.1 \newline $\pm$ 0.06 & 0.044  \\
                            & 25  & 6.45 \newline $\pm$ 0.54 & 91.71  \newline $\pm$ 0.16 & -0.08 \newline $\pm$ 0.06 & 0.089  \\
                            & 50  & 6.55 \newline $\pm$ 0.55 & 91.75  \newline $\pm$ 0.16 & 0 & 0.164 \\

\bottomrule
\end{tabular}
}
\label{tab:nfe}
\end{table}
% %%%%%%%%%%%%%%%%%%%%%%%%%%%%%%%%

\subsection{Analysis on Model Complexities}
\label{sec:model complexity}

To investigate model complexities, we record the number of model parameters and the real-time factor (RTF--the ratio between the model synthesizing time and the duration of the synthesized speech) in Table \ref{tab:model complexity}. RTFs are measured on a single NVIDIA 3090 GPU. As shown in Table \ref{tab:model complexity}, DEX-TTS requires the smallest number of parameters among the expressive TTS methods, showing superior efficiency in the parameter size. However, DEX-TTS has a higher RTF compared to the previous expressive TTS methods. This is a challenge confronted by diffusion-based TTS models, which require multiple denoising steps during speech synthesis (we discuss this further in Sections \ref{sec:nfe and rtf} and \ref{sec:limitation and future work}). On the other hand, GeDEX-TTS achieves more satisfactory results when comparing it with other diffusion-based TTS models in general TTS. GeDEX-TTS achieves the lowest RTF among diffusion-based TTS models with similar parameter sizes. The results demonstrate that our diffusion network design is effective not only for improving performance but also for enhancing inference speed.

\subsection{Analysis on NFE and RTF}
\label{sec:nfe and rtf}

In this subsection, we analyze the proposed models depending on various NFEs. In Table \ref{tab:nfe}, we perform evaluations on the VCTK and LJSpeech datasets for DEX-TTS and GeDEX-TTS, using NFE of 10, 25, and 50. We also conduct a comparative MOS-N (CMOS-N) test to investigate the performance differences in speech naturalness depending on NFEs. The test is performed on 10 participants. The model versions with an NFE of 50 are used as references for comparison. As depicted in Table \ref{tab:nfe}, the naturalness of the synthesized speech slightly decreases as NFE decreases. However, there are no significant performance differences depending on NFEs in objective measures. This indicates that, despite being a diffusion-based TTS, the proposed method can achieve excellent performance even with a small NFE. Specifically, DEX-TTS with an NFE of 10 achieves competitive performance for both TTS performance and efficiency (RTF and parameter size) compared to other expressive TTS methods.

\section{Further Experiments}

%%%%%%%%%%%%%%%%%%%%%%%%%%%%%%%
\begin{table}[t]
\caption{Experiment results on the VCTK and ESD dataset depending on the patch size.}
\centering
\resizebox{0.8\linewidth}{!}{%
\begin{tabular}{l|l|cccc}
\toprule
\multirow{2.5}{*}{Dataset} & \multirow{2.5}{*}{Model} & \multicolumn{2}{c}{Seen scenarios} & \multicolumn{2}{c}{Unseen scenarios} \\ 
\cmidrule{3-6} & & WER & COS & WER & COS \\

\midrule

\multirow{3}{*}{VCTK}   & {DEX-TTS-P2 \hspace{0pt plus 1filll}} 
                           & 7.85 \ $\pm$ 0.73 & 85.31 \ $\pm$ 0.22 & 5.84 \ $\pm$ 0.61 & 80.45 \ $\pm$ 0.26  \\
                        & {DEX-TTS-P4 \hspace{0pt plus 1filll}} 
                           & 6.82 \ $\pm$ 0.70 & 84.19 \ $\pm$ 0.23 & 6.05 \ $\pm$ 0.66 & 79.27 \ $\pm$ 0.27 \\
                        & {DEX-TTS-P8 \hspace{0pt plus 1filll}} 
                           & 6.93 \ $\pm$ 0.69 & 83.94 \ $\pm$ 0.23 & 5.95 \ $\pm$ 0.64 & 78.47 \ $\pm$ 0.28  \\

\midrule

\multirow{3}{*}{ESD}    & {DEX-TTS-P2 \hspace{0pt plus 1filll}} 
                           & 8.34  \ $\pm$ 0.95 & 82.71 \ $\pm$ 0.30 & 8.35 \ $\pm$ 0.93 & 75.58 \ $\pm$ 0.31  \\
                        & {DEX-TTS-P4 \hspace{0pt plus 1filll}} 
                           & 9.30  \  $\pm$ 1.01 & 82.60 \ $\pm$ 0.30 & 9.00 \ $\pm$ 1.01 & 73.02 \ $\pm$ 0.40 \\
                        & {DEX-TTS-P8 \hspace{0pt plus 1filll}} 
                           & 10.09 \ $\pm$ 1.10 & 81.92 \ $\pm$ 0.29 & 9.60 \ $\pm$ 1.03 & 74.04 \ $\pm$ 0.31  \\

\bottomrule
\end{tabular}
}
\label{tab:patch size}
\end{table}
%%%%%%%%%%%%%%%%%%%%%%%%%%%%%%%

In this section, we conduct additional experiments not covered in the main text. These experiments include adjusting the patch size and setting up zero-shot scenarios using unseen emotions to analyze the proposed methods.

\subsection{Experiments depending on Patch Size}

To analyze the model performance depending on the patch size $P$ of our networks, we conduct the experiments in Table \ref{tab:patch size}. We utilize $P$ of 2, 4, and 8 for the experiments. As shown in Table \ref{tab:patch size}, we observe an improvement in WER in the seen scenarios of the VCTK dataset when we use patch sizes larger than 2. However, performance degrades for the other metrics and an overall performance decrease is observed for the ESD dataset. In particular, when comparing the COS performance between patch sizes 2 and 8, performance degradation is evident with a patch size of 8 in both datasets. It indicates that DiT blocks can extract detailed representations of patches when smaller patches are used.

\subsection{Experiments on the Unseen Emotion}

Whereas we set zero-shot scenarios with unseen speakers for the ESD dataset in the main text, we design three zero-shot scenarios based on a few combinations of emotions in this subsection. As depicted in Table \ref{tab:unseen-emotion esd}, the emotion columns indicate the emotion lists for each scenario. That is, the emotion list in the seen scenarios is used for training DEX-TTS. We observe that WER is similar between the emotion zero-shot and speaker zero-shot experiments (Table \ref{tab:exp-esd}--average WER for seen and unseen scenarios is 8.34). However, in the speaker zero-shot experiment, a notable difference of 7.13 is observed in the COS performance between the seen and unseen scenarios, whereas in the emotion zero-shot experiment, the COS performance difference is only approximately 3. This result indicates that adapting to unseen speakers is more difficult than adapting to unseen emotions. In addition, the consistent performance across various emotion zero-shot scenarios suggests that DEX-TTS can adapt to diverse unseen emotions.

\subsection{Experiments on the VCTK dataset for GeDEX-TTS}

In the main text, we conducted experiments on the single-speaker dataset, the LJSpeech dataset, to verify our improved diffusion-based TTS model, GeDEX-TTS. We further perform experiments on the multi-speaker dataset to investigate the results. We use the VCTK dataset as the multi-speaker dataset. To enable GeDEX-TTS to synthesize speech depending on pre-defined speakers, we utilize a common technique, speaker embeddings using a lookup table, as in \cite{popov2021grad}. Table \ref{tab:ge-vctk} shows the comparison results on the VCTK dataset. MOS-N is recorded by evaluations of 16 participants. We observe that GeDEX-TTS outperforms the previous methods, indicating GeDEX-TTS can also synthesize high-quality speech in multi-speaker settings.

\subsection{Further Comparison Results in General TTS}

In the main text, we did not compare GeDEX-TTS with U-DiT-TTS since an official code is not provided. Instead, we verified our strategies by conducting experiments using various encoding types (Table \ref{tab:ge-ab-ljspeech}) and patch sizes (Table \ref{tab:patch size}). To further investigate the effect of our strategies, we reproduce U-DiT-TTS based on the GeDEX-TTS system. We remove the patchify and embedding strategies and utilize the large patch sizes mentioned in their study. As presented in Table \ref{tab:ge-udit}, GeDEX-TTS outperforms our implemented U-DiT-TTS in both WER and COS across various datasets. The experimental results validate that leveraging small patch size, overlapping patchify, and conv-freq embedding strategies enables the effective utilization of DiT block.

%%%%%%%%%%%%%%%%%%%%%%%%%%%%%%%
\begin{table}
\caption{Evaluation for unseen emotions on the ESD dataset.}\label{tab:unseen-emotion esd}
\centering
\resizebox{0.8\columnwidth}{!}{%
\begin{tabular}{p{3cm}|M{1.2cm}M{1.2cm}M{1.2cm}cM{1.2cm}M{1.2cm}M{1.2cm}}
% \begin{tabular}{p{2.5cm}ccccccccccc}
\toprule
\multirow{2.5}{*}{Model} & \multicolumn{3}{c}{Seen scenarios} & & \multicolumn{3}{c}{Unseen (zero-shot) scenarios} \\ 
\cmidrule{2-4} \cmidrule{6-8} & Emotion & WER  & COS  & & Emotion &  WER & COS   \\
\midrule
DEX-TTS     & Happy, Neutral, Surprise & 9.23  \newline $\pm$ 0.99 & 82.05 \newline $\pm$ 0.30 
            & & Angry, Sad             & 7.71  \newline $\pm$ 0.87 & 80.00 \newline $\pm$ 0.33 \\
\midrule
DEX-TTS     & Happy, Sad, Surprise     & 9.28  \newline $\pm$ 0.94 & 82.51 \newline $\pm$ 0.31
            & & Angry, Neutral         & 7.83  \newline $\pm$ 0.90 & 78.85 \newline $\pm$ 0.36 \\
\midrule
DEX-TTS     & Angry, Neutral, Sad      & 7.44  \newline $\pm$ 0.89 & 82.80 \newline $\pm$ 0.30 
            & & Happy, Surprise        & 8.53  \newline $\pm$ 0.87 & 79.43 \newline $\pm$ 0.33 \\
\bottomrule
\end{tabular}}
\end{table}
%%%%%%%%%%%%%%%%%%%%%%%%%%%%%%%

%%%%%%%%%%%%%%%%%%%%%%%%%%%%%%%
\begin{table}[t]
\caption{Results on the VCTK dataset for general TTS task.}
\centering
\resizebox{0.55\linewidth}{!}{%
\begin{tabular}{l|cccc}
\toprule
Model & WER & COS & MOS-N  \\
\midrule
{GT \hspace{0pt plus 1filll}} 
    & 6.23 \ \newline $\pm$ 0.58 & - & 4.38 $\pm$ 0.04  \\
{FastSpeech2 \hspace{0pt plus 1filll}} 
    & 9.80 \ \newline $\pm$ 0.67 & 85.38 \ \newline $\pm$ 0.28 & 3.32 $\pm$ 0.06  \\
{Grad-TTS \hspace{0pt plus 1filll}} 
    & 7.70  \ \newline $\pm$ 1.00 & 85.97  \ \newline $\pm$ 0.31 & 3.91 $\pm$ 0.05  \\
% {UDiT-TTS  \hspace{0pt plus 1filll}} 
%     & 8.91  \ \newline $\pm$ 0.83 & 83.93 \ \newline $\pm$ 0.42 & 3.49 $\pm$ 0.06  \\
{ComoSpeech  \hspace{0pt plus 1filll}} 
    & 14.09 \ \newline $\pm$ 0.88  & 82.66 \ \newline $\pm$ 0.25 & 3.64 $\pm$ 0.06  \\
{GeDEX-TTS \hspace{0pt plus 1filll}} 
    & \textbf{6.36} \ \newline $\pm$ 0.55 & \textbf{86.14} \ \newline $\pm$ 0.19 & \textbf{3.98} $\pm$ 0.05 \\
\bottomrule
\end{tabular}
}
\label{tab:ge-vctk}
\end{table}
%%%%%%%%%%%%%%%%%%%%%%%%%%%%%%%

% %%%%%%%%%%%%%%%%%%%%%%%%%%%%%%%%
\begin{table}[ht]
\caption{Comparison results with U-DiT-TTS.}
\centering
\resizebox{0.5\linewidth}{!}{%
\begin{tabular}{l|l|cc}
\toprule
Dataset & Model & WER & COS \\
\midrule
\multirow{2}{*}{LJSpeech} & U-DiT-TTS   & 8.91 \ \newline $\pm$ 0.61 & 86.76 \ \newline $\pm$ 0.20 \\
                          & GeDEX-TTS  & 6.55 \ \newline $\pm$ 0.55 & 91.75 \ \newline $\pm$ 0.16 \\
\midrule
\multirow{2}{*}{VCTK}     & U-DiT-TTS   & 8.91 \ \newline $\pm$ 0.83 & 83.93 \ \newline $\pm$ 0.42 \\
                          & GeDEX-TTS  & 6.36 \ \newline $\pm$ 0.55 & 86.14 \ \newline $\pm$ 0.19 \\
\bottomrule
\end{tabular}
}
\label{tab:ge-udit}
\end{table}
% %%%%%%%%%%%%%%%%%%%%%%%%%%%%%%%%

% \subsection{Experiments on a larger dataset}
% LibriTTS

\section{Visualizations}

\subsection{Style Visualizations}
\label{sec:style viz}

To further investigate the proposed model, we visualize the extracted T-IV and T-V styles. Based on the T-IV and T-V encoders, we first extract each style ($h_{inv}$, $h^{e}_{v}$, and $h^{d}_{v}$) from the reference speech. Since $h^{d}_{v}$ has the time dimension, we apply a channel-wise average to $h^{d}_{v}$ to obtain a style vector. Then, we utilize T-SNE to visualize each style. We include $h_{inv}+h^{e}_{v}$ to analyze the synergy between $h_{inv}$ and $h^{e}_{v}$. We obtain $h_{inv}+h^{e}_{v}$ by concatenating $h_{inv}$ and $h^{e}_{v}$. In addition, we record the distance within the cluster (DWC) and the distance between clusters (DBC) to analyze deeper.

\begin{figure*}[t]
  \centering
  \includegraphics[scale=0.45]{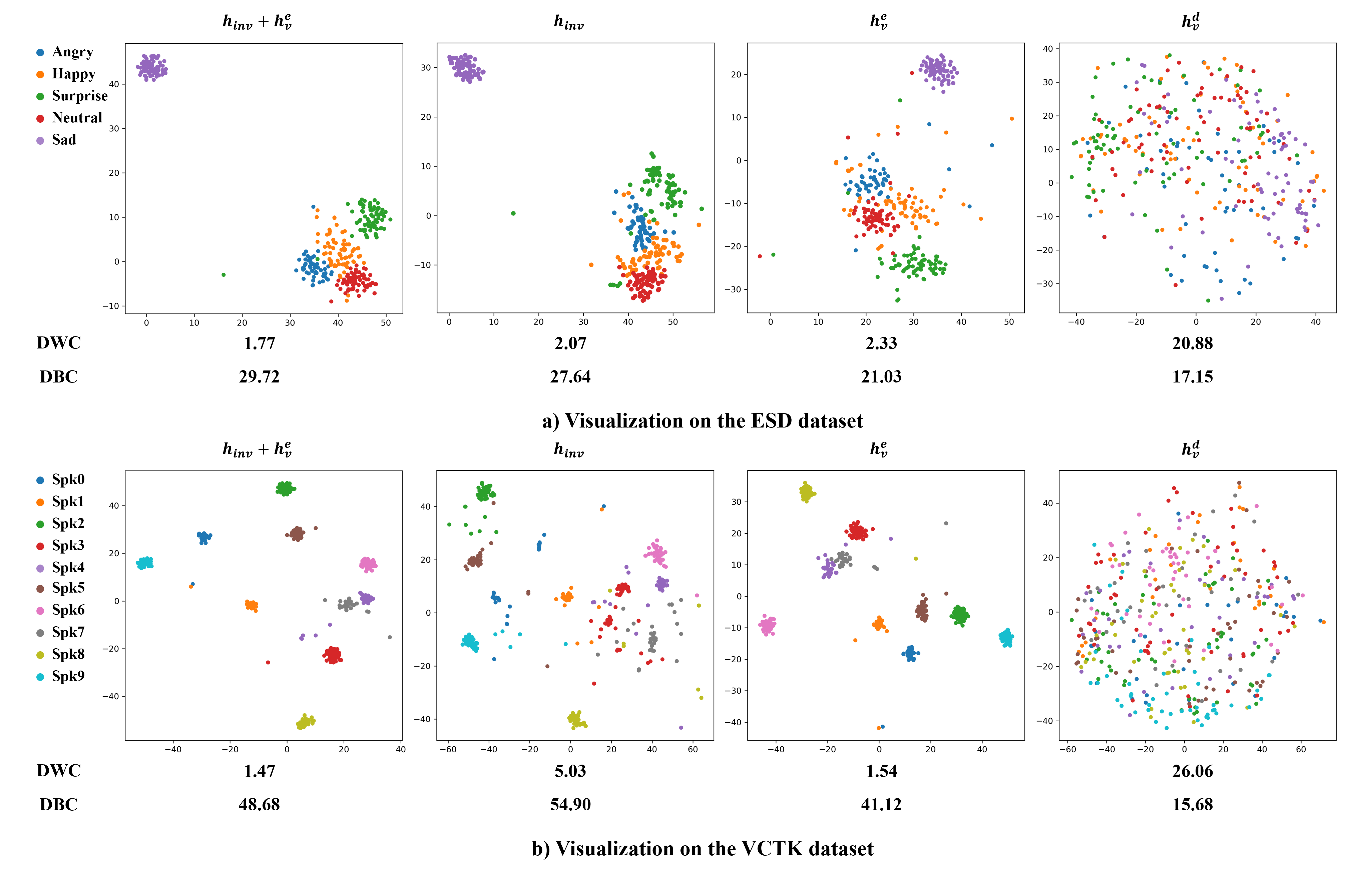} 
  \caption{Style visualizations using T-SNE on the ESD and VCTK datasets. DWC and DBC indicate distance within clusters and distance between clusters. For the ESD dataset, T-SNE is used based on the five emotions of speaker 0016. For the VCTK dataset, T-SNE is applied based on unseen speakers of the dataset. DEX-TTS trained with each dataset is used for style extraction.} 
  \label{fig:sty_viz}
\end{figure*}

As shown in Figure \ref{fig:sty_viz}.a), we visualize each style depending on the emotions in the ESD dataset. $h_{inv}$ and $h^{e}_{v}$ show dense clusters based on emotions, indicating that both time-invariant and time-variant styles encompass emotion-related information. Furthermore, $h_{inv}+h^{e}_{v}$ provides superior clustering results with a decrease in DWC and an increase in DBC compared to $h_{inv}$ and $h^{e}_{v}$. It suggests that $h_{inv}$ and $h^{e}_{v}$ represent distinct information and that the synergy arises when both styles are utilized. These results align with the findings of the ablation studies (Table \ref{tab:ab-esd}), proving the necessity of both $h_{inv}$ and $h^{e}_{v}$. The results for the VCTK dataset are shown in Figure \ref{fig:sty_viz}.b). We observe similar results to those of the ESD dataset. $h_{inv}$ and $h^{e}_{v}$ form clusters based on speakers, while $h_{inv}+h^{e}_{v}$ yields better clustering results than $h_{inv}$ and $h^{e}_{v}$ through synergy. The consistent visualization results across various datasets verify that the proposed style modeling can capture emotion or speaker information without explicit labels.

Lastly, we find interesting results for $h^{d}_{v}$ in both datasets. Unlike $h_{inv}$ and $h^{e}_{v}$, $h^{d}_{v}$ does not form clusters based on emotions or speakers. However, considering the effect on the performance of $h^{d}_{v}$ in ablation studies (Table \ref{tab:ab-esd}), $h^{d}_{v}$ contains significant time-variant styles besides, speaker or emotions. It suggests that even the same types of time-variant styles ($h^{e}_{v}$ and $h^{d}_{v}$) can contain different information depending on the extraction and reflection methods. In summary, the proposed style modeling method can extract diverse styles including speaker and emotional information, achieving well-represented styles for expressive TTS.

\subsection{Mel-Spectrograms Visualizations}

\begin{figure*}[t]
  \centering
  \includegraphics[scale=0.65]{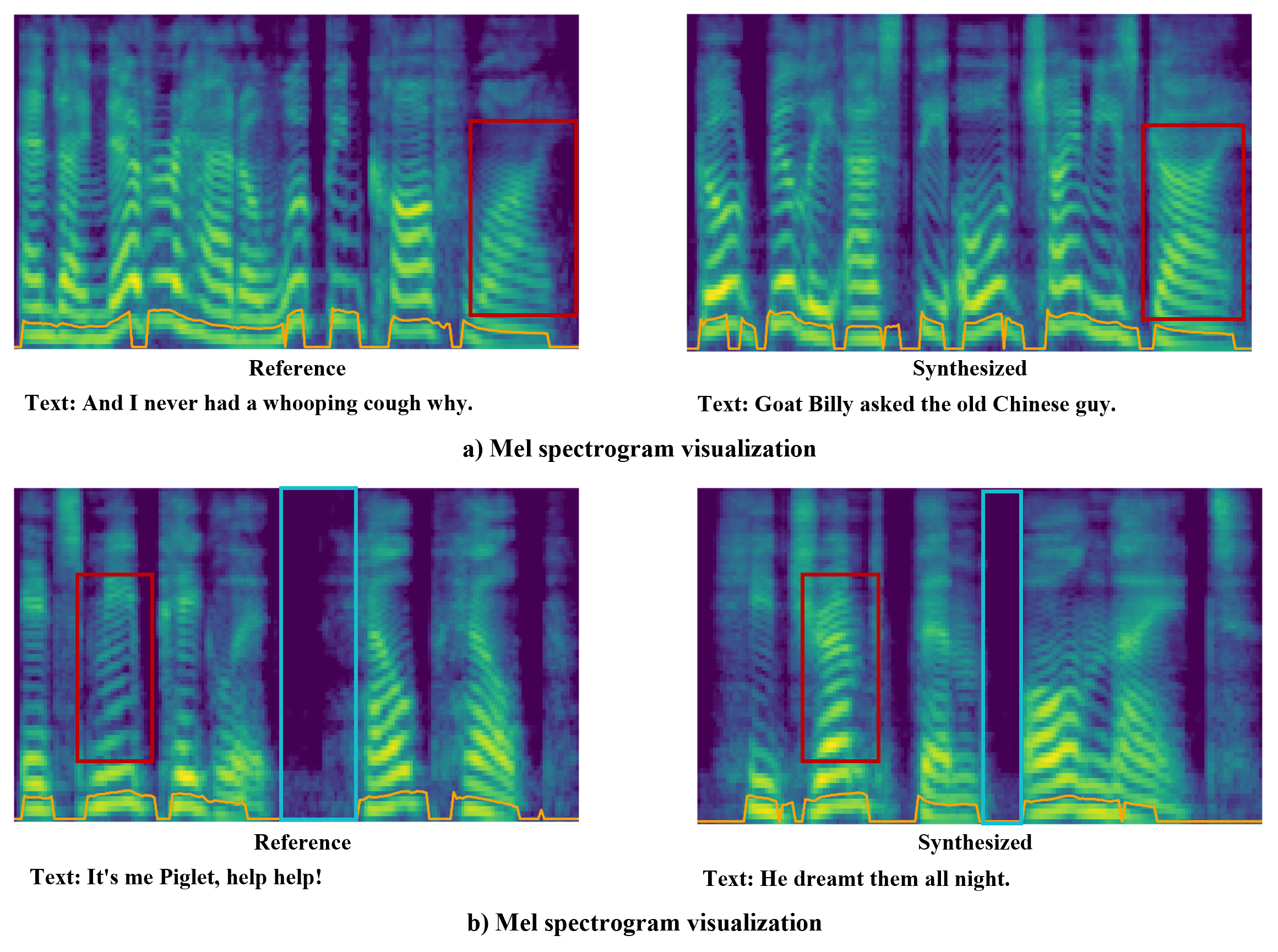} 
  \caption{Visualization of mel-spectrograms for reference and synthesized speech on the ESD dataset. The orange lines indicate pitch information. Red boxes are used for comparing frequency bins and blue boxes are used for comparing pause points style.} 
  \label{fig:mel_viz}
\end{figure*}

In this subsection, we plot mel-spectrograms and pitch for non-parallel samples of the ESD dataset. As shown in Figure \ref{fig:mel_viz}, synthesized speech represents diverse styles of reference speech. In specific, DEX-TTS can follow the prosodic styles of the reference and properly represent it for a given text (see the orange lines in Figure \ref{fig:mel_viz}). The red boxes in the plots indicate that DEX-TTS can make detailed frequency bins, similar to those of reference speech. In addition, we observe that DEX-TTS can resemble other reference styles, beyond the prosodic or detailed frequency styles. The blue boxes in Figure \ref{fig:mel_viz}.b) demonstrate that the synthesized speech contains the intermediate pause point like reference speech. To understand the results, we provide demos of these samples at our demo site.

\section{Limitation and Future Work}
\label{sec:limitation and future work}

As discussed in Sections \ref{sec:model complexity} and \ref{sec:nfe and rtf}, diffusion-based TTS models generally exhibit higher RTF than non-diffusion-based TTS models since they require an iterative denoising process. Nevertheless, the proposed methods are efficient in terms of parameter size and exhibit faster inference speed compared to other diffusion-based TTS models. In addition, the proposed methods can achieve competitive performance even with fewer NFEs. Despite inspiring results with competitive RTFs, the iterative denoising process inherent in diffusion-based TTS remains a challenge.

Recent studies \cite{song2023consistency, kim2023consistency} have provided suggestions to address the limitations of this study. Song et al. \cite{song2023consistency} introduced a consistency model (CM) that can map any point $x_{t}$ on the ODE trajectory to its origin $x_{0}$ for generative modeling. Since the model is trained to be consistent for points on the same trajectory, these models are referred to as consistency models. By adopting distillation methods, they obtained CM and generated high-quality data with only one sampling process. CoMoSpeech \cite{ye2023comospeech} also adopted CM to accelerate the sampling process in diffusion-based TTS. However, as the consistency trajectory model (CTM) \cite{kim2023consistency} mentioned, CM does not exhibit a speed-quality trade-off (i.e., the generation quality does not improve as NFE increases). They introduced an alternative way to bridge score-based and distillation models to accelerate the sampling process while maintaining a speed-quality trade-off. Inspired by their works, we plan to extend our study to accelerate the sampling process for the proposed diffusion-based TTS models. In future work, we will address the limitations of this study and propose diffusion-based TTS models with excellent performance and fast generation capabilities while maintaining the speed-quality trade-off.

\section{Subjective evaluation}
\label{sec:subjective evaluation}

\begin{figure*}[ht]
  \centering
  \includegraphics[scale=0.6]{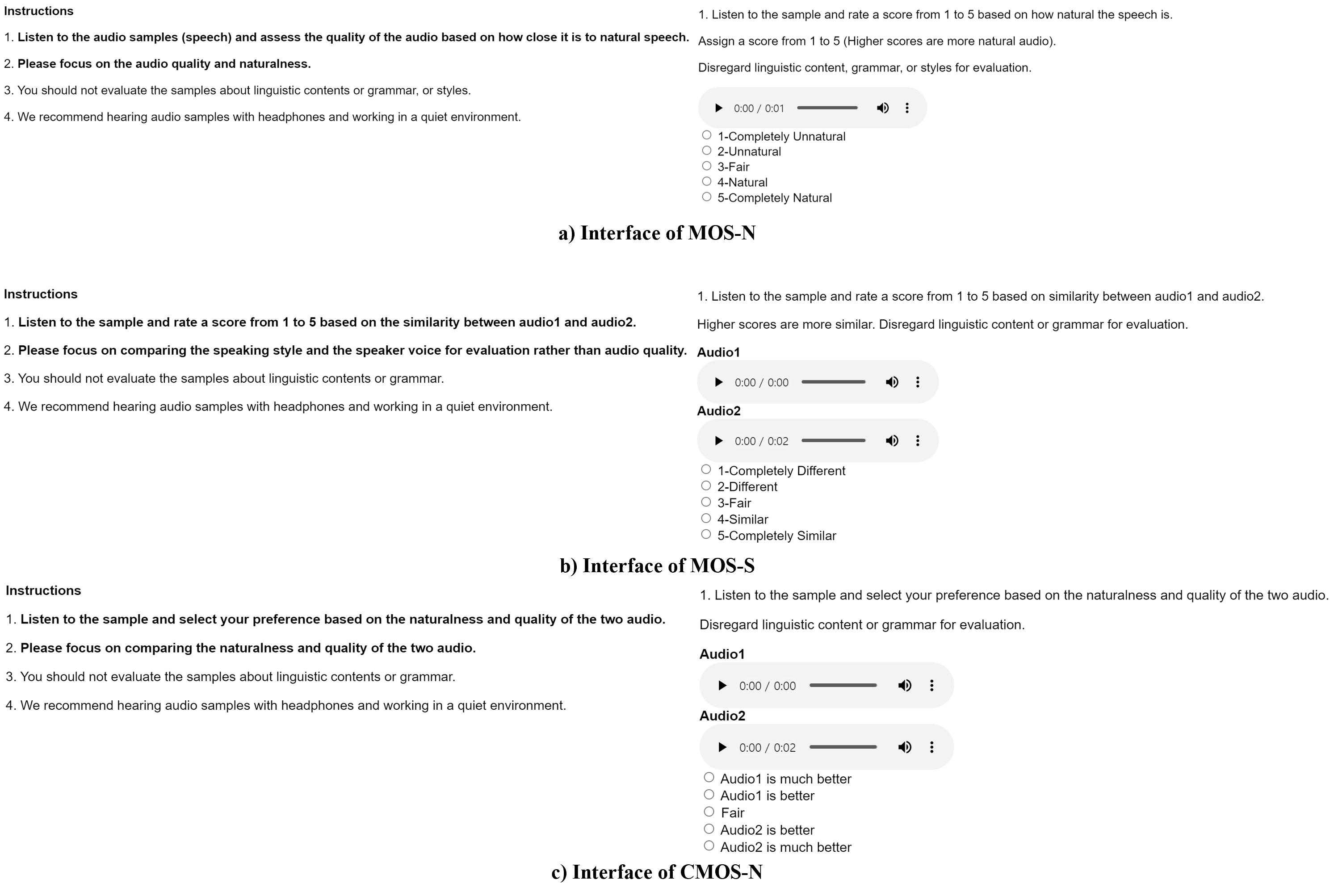} 
  \caption{Interfaces of MOS evaluations.} 
  \label{fig:mos_interface}
\end{figure*}

We provide MOS evaluation interface (screenshot) in Figure \ref{fig:mos_interface}. It contains an overall interface and instructions for participants.

\end{document}